\documentclass[twocolumn]{aastex63}

\usepackage{
	amsmath,
	amssymb,
	amsthm,
	microtype,
    stmaryrd,
    xcolor, 
}
\usepackage{savesym}
\savesymbol{tablenum}
\usepackage{siunitx}
\usepackage{rotating}
\usepackage{xspace}
\restoresymbol{SIX}{tablenum}
\DeclareSIUnit{\jy}{Jy}
\DeclareSIUnit{\beam}{beam}
\newcommand{\kms}{\ensuremath{\mathrm{km~s}^{-1}\xspace}}

\usepackage[version=4]{mhchem}

\usepackage{xcolor}

\DeclareUnicodeCharacter{2212}{-}

\shorttitle{G5 Cloud-Cloud Collision}
\shortauthors{Gramze et al.}

\graphicspath{{./}{figures/}}

\begin{document}
\newpage

\title{Evidence of a Cloud-Cloud Collision from Overshooting Gas in the Galactic Center}

\correspondingauthor{Savannah R. Gramze}
\email{savannahgramze@ufl.edu}

\author[0000-0002-1313-429X]{Savannah R. Gramze}
\affiliation{Department of Astronomy, University of Florida, Gainesville, FL 32611 USA}

\author[0000-0001-6431-9633]{Adam Ginsburg}
\affiliation{Department of Astronomy, University of Florida, Gainesville, FL 32611 USA}

\author[0000-0001-9436-9471]{David S. Meier}
\altaffiliation{}
\affiliation{New Mexico Institute of Mining and Technology, 801 Leroy Pl, Socorro, NM 87801 USA}
\affiliation{National Radio Astronomy Observatory, PO Box O, Socorro, NM 87801 USA}

\author[0000-0001-8224-1956]{Juergen Ott}
\altaffiliation{}
\affiliation{National Radio Astronomy Observatory, PO Box O, Socorro, NM 87801 USA}

\author[0000-0002-0133-8973]{Yancy Shirley}
\affiliation{Steward Observatory, University of Arizona, 933 North Cherry Avenue, Tucson, AZ 85721}

\author[0000-0001-6113-6241]{Mattia C. Sormani}
\affiliation{Department of Physics, University of Surrey, Guildford GU2 7XH, UK}
\affiliation{Universität Heidelberg, Zentrum für Astronomie, Institut für theoretische Astrophysik, Albert-Ueberle-Str. 2, 69120 Heidelberg, Germany}

\author[0000-0002-8502-6431]{Brian E.\ Svoboda}
\affiliation{National Radio Astronomy Observatory, PO Box O, Socorro, NM 87801 USA}

\begin{abstract}

The Milky Way is a barred spiral galaxy with ``bar lanes" that bring gas towards the Galactic Center. Gas flowing along these bar lanes often overshoots, and instead of accreting onto the Central Molecular Zone, it collides with the bar lane on the opposite side of the Galaxy. We observed G5, a cloud which we believe is the site of one such collision, near the Galactic Center at $(\ell,b) = (+5.4, -0.4)$ with the ALMA/ACA. We took measurements of the spectral lines \ce{^12CO} $J=2\shortrightarrow1$
, \ce{^13CO} $J=2\shortrightarrow1$
, \ce{C^18O} $J=2\shortrightarrow1$,
\ce{H2CO} $J=3_{03}\shortrightarrow2_{02}$
, \ce{H2CO} $J=3_{22}\shortrightarrow2_{21}$
, \ce{CH3OH} $J=4_{22}\shortrightarrow3_{12}$
, \ce{OCS} $J=18\shortrightarrow17$ and \ce{SiO} $J=5\shortrightarrow4$. 
We observed a velocity bridge between two clouds at $\sim$\SI{50}{\kms} and $\sim$\SI{150}{\kms} in our position-velocity diagram, which is direct evidence of a cloud-cloud collision. 
We measured an average gas temperature of $\sim$\SI{60}{K} in G5 using \ce{H2CO} integrated intensity line ratios. 
We observed that the \ce{^12C}/\ce{^13C} ratio in G5 is consistent with optically thin, or at most marginally optically thick \ce{^12CO}. 
We measured \SI{1.5e19}{cm^{-2}~(K~\kms)^{-1}} for the local X$_{\rm CO}$, 10-20x less than the average Galactic value. 
G5 is strong direct observational evidence of gas overshooting the Central Molecular Zone (CMZ) and colliding with a bar lane on the opposite side of the Galactic center. 

\end{abstract}

\keywords{
    ISM: clouds ---
    ISM: molecules ---
    ISM: structure ---
    Galactic Center ---
    astrochemistry
}

\section{Introduction} 

\begin{figure*}[t]
    \centering
    \includegraphics[width=\textwidth]{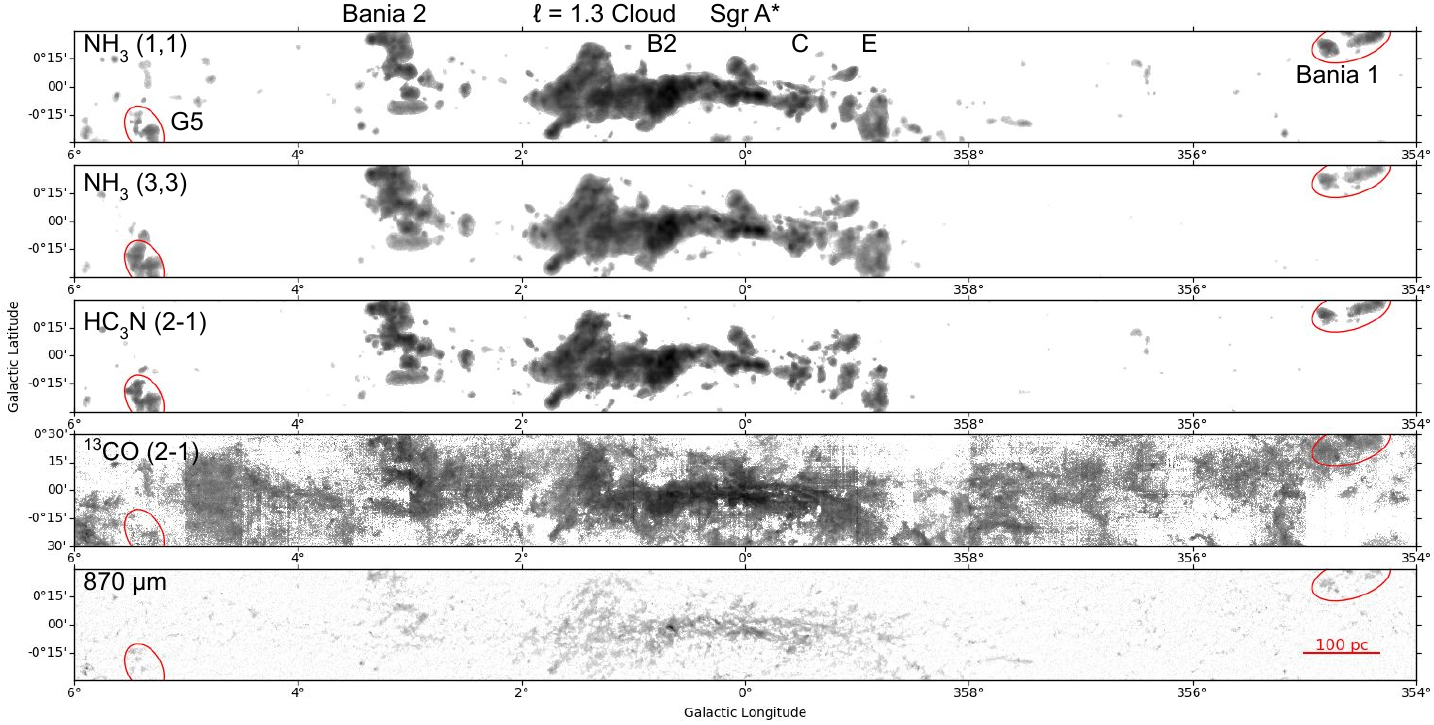}
    \caption{View of the inner \SI{12}{\degree} of the Galactic plane. G5 at $(\ell,b) = (+5.4, -0.4)$ and Bania 1 at $(\ell,b) = (-5.4, +0.4)$ are circled in red. 
    The first three panels are \ce{NH3} (1,1), \ce{NH3} (3,3) and \ce{HC3N} (2-1) from the Mopra HOPS survey \citep{hops3, hops1, hops2}.
    The fourth panel is \ce{^13CO} (2-1) from the APEX SEDIGISM survey \citep{SEDIGISM2021}. 
    The last panel is \SI{850}{\micron} from the ATLASGAL survey \citep{atlasgal2009}.
    }
    \label{fig:galoverview}
\end{figure*}

\begin{figure}[t]
    \centering
    \includegraphics[width=.25\textwidth]{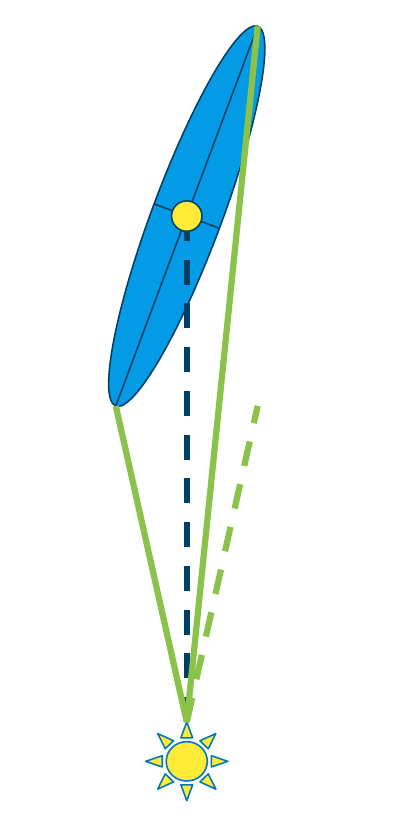}
    \caption{Geometry of the line-of-sight looking at the Milky Way's Galactic bar, depicted as the blue ellipse bisected along its major and minor semi-axes. Not to scale. The dark blue dotted line is the line-of-sight to the center of the Galaxy, with the CMZ as the yellow ellipse. The solid green lines are lines-of-sight to the ends of the Galactic bar. The dotted green line is the same angle from the center of the Galaxy line as the solid green line to the closer end of the bar. Going by this geometry, we conclude that features at the same angular distance from the Galactic center are not necessarily related to symmetrical features on the Galactic bar.  \label{fig:geometry}}
\end{figure}

The Milky Way is a barred spiral galaxy \citep{blitz1991,wegg2013}. 
It has a central, tri-axial bar. The major axis extends out to Galactocentric radii $R\sim$ \SI{5}{kpc} and forms an angle with the Sun-Galactic center line of \SI{20}-\SI{35}{\degree} \citep{blandhawthorn2016}. 
The bar generates a strongly non-axisymmetric gravitational potential, resulting in non-circular stellar and gaseous orbits. 

At the center of the bar is the Central Molecular Zone (CMZ), host to our Galaxy's supermassive blackhole. Figure \ref{fig:galoverview} shows the inner \SI{12}{\degree} of the Galactic plane in various surveys, with the CMZ at the center.
The major axis of the bar is inclined relative to our line of sight, see the top down view in Figure \ref{fig:geometry}, so that the near (far) end of the bar lies at positive (negative) Galactic longitudes. 

The dynamics of gas in the Galactic bar can be broadly understood by considering closed periodic orbits. 
The two most important classes of orbits in a barred potential are X$_1$ and X$_2$ orbits \citep{contopoulos1989}. 
X$_1$ orbits are elongated with their longest axis aligned with the major axis of the Galactic bar, and can become self-intersecting at smaller radii with small cusps at both ends.
X$_2$ orbits are ensconced within X$_1$ orbits at the very center of the bar. The CMZ is believed to be made of gas, dust and stars moving along X$_2$ orbits at the center of the Galaxy. 

Gas flows along the elongated X$_1$ orbits while slowing drifting inwards as it loses angular momentum. 
At some point, the inner X$_1$ orbits become self-intersecting, and gas can no longer follow them, leading to the formation of large-scale shocks as the gas plunges within a dynamical time to X$_2$ orbits that lie much closer to the center of the Galaxy. 
The large-scale shocks associated with the transition between X$_1$ and X$_2$ orbits observationally correspond to the bar lanes which are observed in barred galaxies such as NGC 1300 and NGC 6782. In the Milky Way, these bar lanes have been identified in CO data cubes \citep{fux1999, marshall2008}. 

Gas flows along the Galactic bar lanes at a rate that has been estimated to be \SI{2.6}{M_{\odot} yr^{-1}} \citep{sormanibarnes19}.
Only about a third of this gas accretes onto the CMZ, at a rate of \SI{0.8}{M_{\odot} yr^{-1}} \citep{Hatchfield2021}. Note that these values are obtained assuming a Galactic-averaged X$_{\rm CO}$ factor, but as we shall see below the latter might be significantly lower in the bar lanes, leading to a lower accretion rate.
The accreted gas fuels star formation in the CMZ, which is currently occuring at a rate of \SI{0.04}{}-\SI{0.1}{M_{\odot} yr^{-1}} \citep[][]{Yusef-Zadeh2009, Immer2012, longmore2013}. 
The gas not accreted onto the CMZ overshoots the CMZ, eventually colliding with the bar lane on the other side of the Galaxy.

To better understand the Galactic bar, we view the inner \SI{12}{^{\circ}} of the Galactic Center in Figure \ref{fig:galoverview}. Immediately noticeable in \ce{NH3} (3,3) emission outside of the CMZ are two clouds at $(\ell,b) = (+5.4, -0.4)$ and $(\ell,b) = (-5.4, +0.4)$. We call G5 the cloud at $(\ell,b) = (+5.4, -0.4)$. The cloud at $(\ell,b) = (-5.4, +0.4)$ is identified as Bania 1 (B1) \citep{bania86}. 
These two clouds, G5 and B1, are remarkable in that they are the furthest large regions of bright \ce{NH3} (3,3) emission outside of the CMZ. 

In this paper, we investigate the G5 cloud along the Galactic bar. We present two mosaicked fields of molecular line observations from ALMA/ACA using the Total Power array. We investigate the spectral components and velocity structure of G5. Next, we measure the cloud's gas temperature using \ce{H2CO} molecular lines. Then, we estimate a portion of the cloud's mass by comparing various mass estimation methods. Finally, we discuss the properties of G5 and their implications for the cloud's positions on the Galactic bar.

\section{Observations}

\begin{figure*}[ht]
    \centering
    \includegraphics[width=0.7\textwidth]{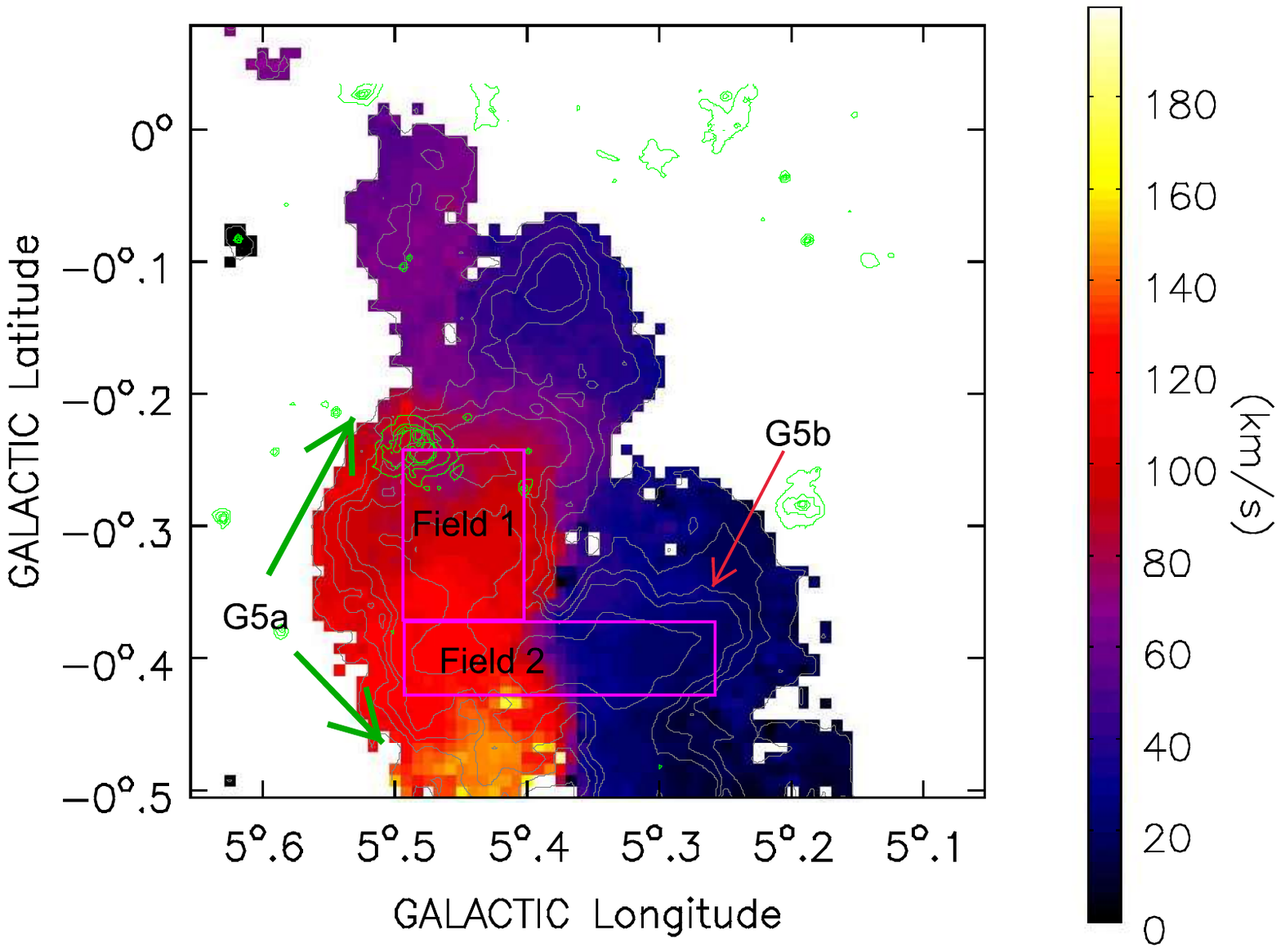}
    \caption{Velocity field (colours) and integrated intensity contours (grey) from the NH$_3$ (3,3) Mopra HOPS Survey \citep{hops3, hops1, hops2}. The contour levels of the grey moment 0 are [0.2, 0.4, 0.6, 0.8, 1.5]. Green contours are from 70 micron Hi-GAL, with contour levels of [0.1, 0.2, 0.4, 0.6, 0.8] \citep{higal2016}. The magenta boxes outline the fields. The clouds G5a and G5b are labeled. 
    \label{fig:g5map}}
\end{figure*}

\begin{figure}
    \centering
    \includegraphics[width=0.45\textwidth]{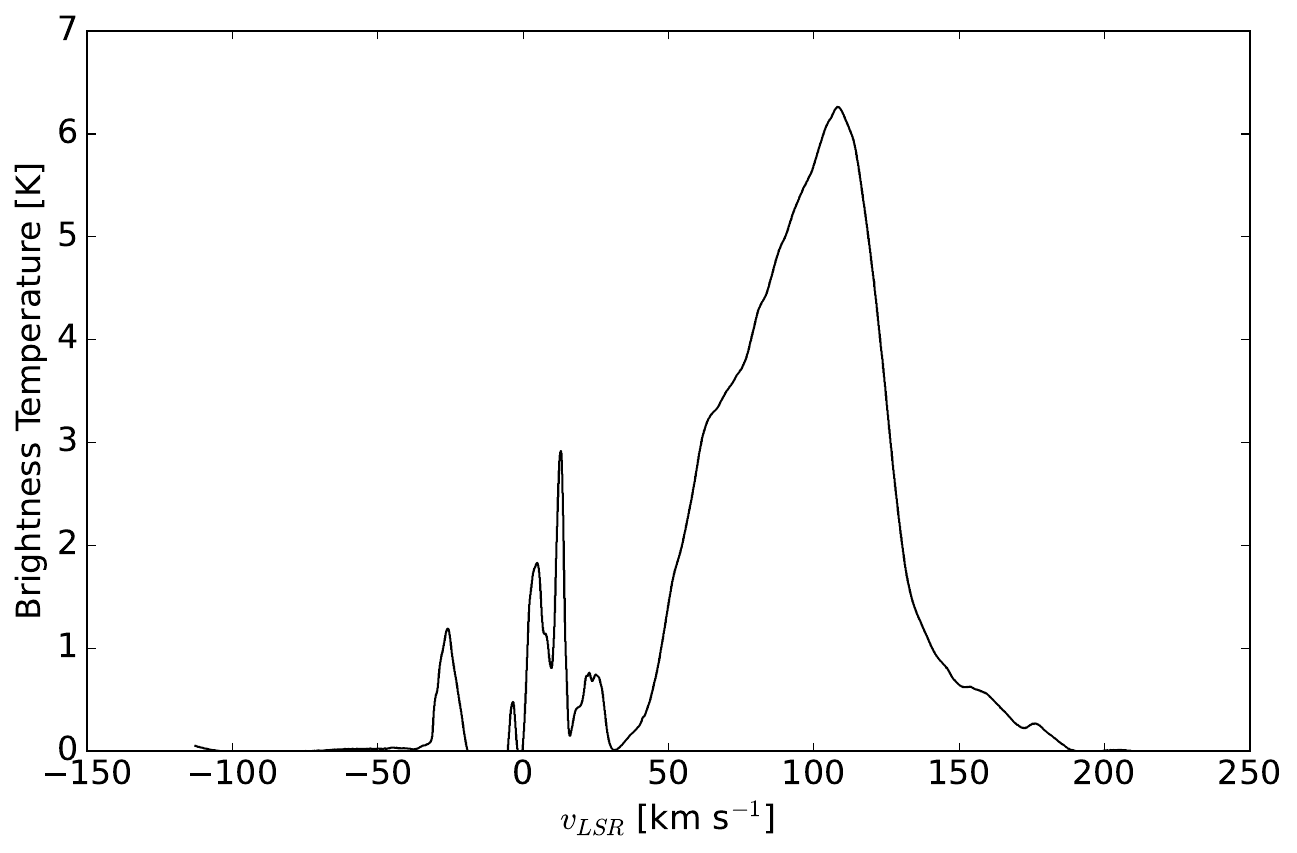}
    \includegraphics[width=0.45\textwidth]{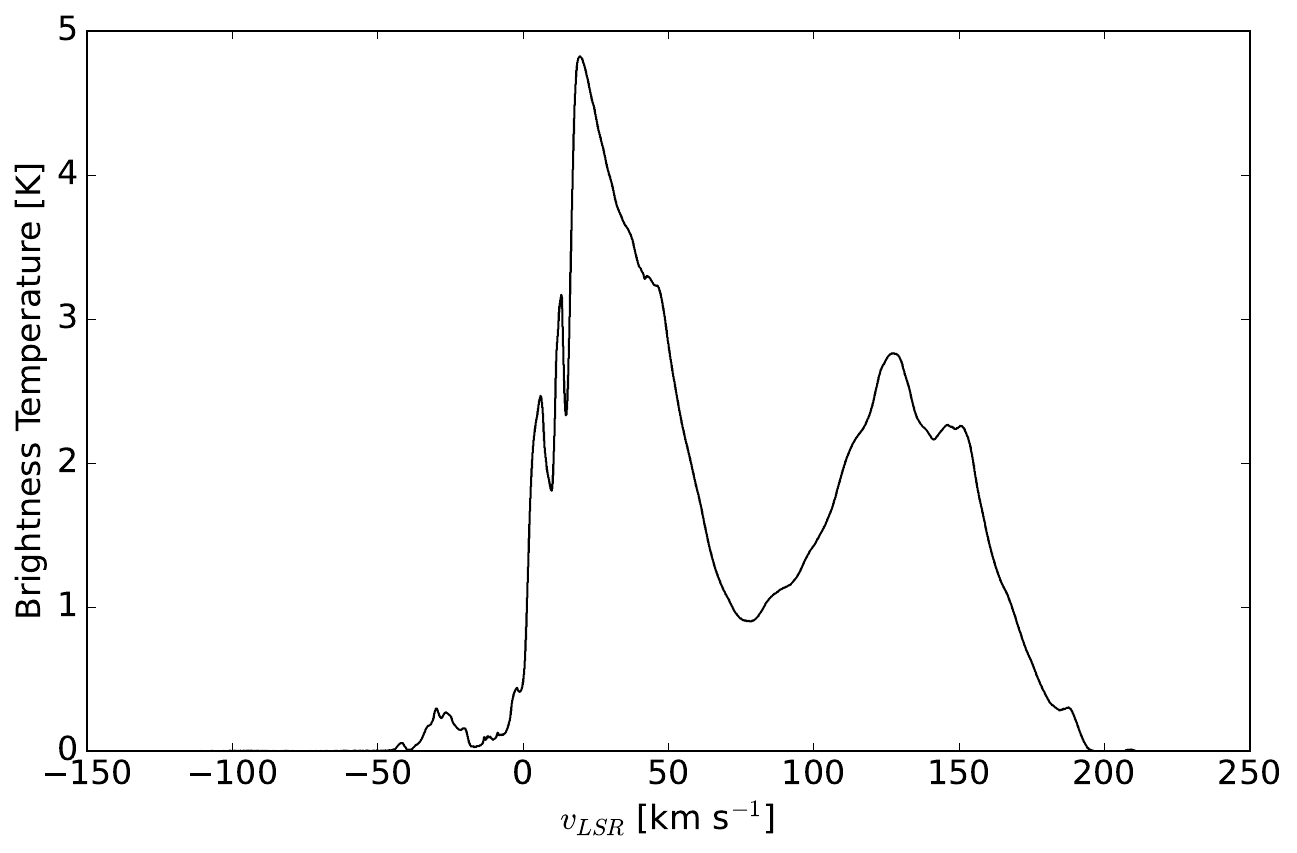}
    \caption{Averaged spectra of \ce{^12CO} $2\shortrightarrow1$ over the two fields defined in Figure \ref{fig:g5map}. 
    Top: Field 1, which contains emission from cloud G5a from velocities \SI{50}{\kms} to \SI{200}{\kms} that continues down into Field 2 at a higher velocity, and also a line of sight HII region disassociated with G5 at a lower velocity. 
    Bottom: Field 2, which contains emission from both clouds G5a and G5b. 
    We identify G5a as associated with the emission from \SI{75}{\kms} to \SI{200}{\kms}, and G5b as the emission from \SI{15}{\kms} to \SI{75}{\kms}.
    G5b likely continues in emission to $\sim$ \SI{0}{\kms}, but overlap with line of sight emission from \ce{^12CO} associated with the Galactic disk prevents it from being included. }
    \label{fig:spectra}
\end{figure}

The ALMA Atacama Compact Array (ACA) was used to observe the molecular clouds B1 and G5 (project code 2018.1.00862.S). Both 7m and Total Power observations were made. Only the Total Power observations of G5 are investigated in this paper. 
The Total Power array has a resolution of $\sim$ \SI{30}{^{\prime\prime}} in Band 6 (around 220 GHz), which was used to observe in the pertinent spectral lines. 
We show an overview of G5 as shown in \ce{NH3} (3,3) emission in the Mopra HOPS Survey \citep{hops3, hops1, hops2}. 
We observed G5 with two rectangular fields based on the intensity in \ce{NH3}, one roughly along the middle of the north-south extent of the cloud and one along the east-west. The observed fields are boxed in magenta in Figure \ref{fig:g5map}. The north-south extent we refer to as Field 1 (vertical in Fig \ref{fig:g5map}) and the east-west (horizontal in Fig \ref{fig:g5map}) as Field 2. The two fields overlap each other slightly.
Figure \ref{fig:g5map} shows Field 1 consists of the vertical rectangular region extending over \SI{0.375}{\degree} to \SI{0.225}{\degree} in Galactic latitude and \SI{5.4}{\degree} to \SI{5.5}{\degree} in Galactic longitude. Field 2 in Figure \ref{fig:g5map} is a horizontal region extending over \SI{-0.43}{\degree} to \SI{-0.385}{\degree} in Galactic latitude and \SI{5.48}{\degree} to \SI{5.285}{\degree} in Galactic longitude.
A total of \SI{31}{hours} on the Total Power array were used on the two fields.

\begin{table*}
\centering
\caption{Correlator Configuration}
\label{tab:obs}
\begin{tabular}{cccccccc}
\hline
Molecule and & Center Rest & Einstein & Collision Rates\footnote{From Leiden Atomic and Molecular Database \citep{lambda2005} accessed February 2023} & Critical & Eff.  & Velocity & Velocity \\
and & Frequency & A & ($\mathrm{T\,=\,\SI{60}{K}}$) & Density & Ch. & Bandwidth & Resolution \\
 Transition & $\mathrm{GHz}$ & $\mathrm{s^{-1}\, \times 10^{-6}}$ & $\mathrm{cm^{3}\, s^{-1}\, \times 10^{-11}}$ & $\mathrm{cm^{-3}\, \times 10^5}$ & \# & $\mathrm{km\,s^{-1}}$ & $\mathrm{km\,s^{-1}}$ \\
 \hline
 \hline
\ce{^12CO} $J=2\shortrightarrow1$ & 230.53800000 & 0.691 & 6.0 & 0.115 & 3840 & 304.8 & 0.159 \\
H(30)$\alpha$ & 231.9009278 & - & - & - & 3840 & 2424.2 & 1.263 \\
\hline
\ce{H2CO} $J=3_{2,1}\shortrightarrow2_{2,0}$ & 218.760071 & 254.812 & 9.1 & 28.001 & 960 & 321.2 & 0.774 \\
\ce{OCS} $J=18\shortrightarrow17$ & 218.9033565 & 30.371 & 7.4 & 4.104 & - & - & - \\
\hline
\ce{HC3N} v=0 $J=24\shortrightarrow23$ & 218.32472 & 826.0 & 4.71 & 175.372 & 960 & 321.9 & 0.671 \\
\ce{CH3OH} $J=4_{2,2}\shortrightarrow3_{1,2}$ & 218.440063 & 46.863 & 0.093 & 503.904 & - & - & - \\
\ce{H2CO} $J=3_{2,2}\shortrightarrow2_{2,1}$ & 218.475642 & 253.822 & 9.1 & 27.893 & - & - & - \\
\hline
\ce{H2CO} $J=3_{0,3}\shortrightarrow2_{0,2}$ & 218.222192 & 281.8 & 9.1 & 30.967 & 960 & 322.0 & 0.671 \\
\hline
SiO v=0 $J=5\shortrightarrow4$ & 217.104919 & 519.7 & 20.65 & 25.167 & 960 & 323.7 & 0.674 \\
\hline
\ce{^13CO} $J=2\shortrightarrow1$ & 220.39868420 & 0.604 & 6.0 & 0.101 & 1920 & 318.8 & 0.332 \\
\hline
\ce{C^18O} $J=2\shortrightarrow1$ & 219.56035410 & 0.601 & 6.0 & 0.1 & 1920 & 320.1 & 0.333 \\
\hline
\end{tabular}
\end{table*}

The correlator configuration includes several classes of astronomically important spectral lines simultaneously. The first are the isotopologues of carbon monoxide: \ce{^12CO} $J=2\shortrightarrow1$, \ce{^13CO} $J=2\shortrightarrow1$ and \ce{C^18O} $J=2\shortrightarrow1$.
The second is \ce{SiO} $J=5\shortrightarrow4$, which is a strong shock tracer \citep{schilke97} and should help determine how turbulent these clouds are as a result of shocks. 
Third is HC$_3$N $J=24\shortrightarrow23$ as a dense gas tracer \citep{mills2018}.
Fourth are the two formaldehyde lines, 
\ce{H2CO} $J=3_{03}\shortrightarrow2_{02}$ and 
\ce{H2CO} $J=3_{22}\shortrightarrow2_{21}$, which can be used together as a temperature tracer \citep{mangum1993, ginsburg16}. 
Fifth is the radio recombination line 
H(30)$\alpha$, which traces HII regions. This line is in a sub-band with the widest bandwidth to capture the potentially wide radio recombination line.
The observation parameters of the spectral windows are included in Table \ref{tab:obs}.
For the purposes of analysis, we ignore the spectral lines for \ce{OCS} $J=18\shortrightarrow17$, \ce{CH3OH} $J=4_{22}\shortrightarrow3_{12}$, and \ce{H2CO} $J=3_{22}\shortrightarrow2_{21}$.

\section{Data Reduction}

\subsection{Target Description}

G5 is a giant molecular cloud at $(\ell,b) = (+5.4, -0.4)$. 
The result of \citet{GravityCollab19} was that the distance to the Galactic Center is \SI{8178}{pc} with an error of 0.3\%.
The angle between the major axis of the Galactic bar and the Sun-GC line is less certain, but we will assume it to be $\sim$ \SI{30}{\degree} $\pm$ \SI{2}{\degree} \citep{ wegg2015, blandhawthorn2016}.
Assuming that G5 is located on the Galactic bar, and that the distance is uncertain by $\sim$ \SI{1}{kpc} in the direction perpendicular to the bar major axis due to the finite width of the bar, we find using the Law of Sines that G5 is at a distance of \SI{7.06}{kpc} $\pm$ \SI{0.88}{kpc} from the Sun. The distance between G5 and the GC is \SI{1.33}{kpc} $\pm$ \SI{0.15}{kpc}.

G5 has a projected extent of \SI{48.9}{pc} in Galactic latitude and \SI{42.8}{pc} long in Galactic longitude at a distance of \SI{7}{kpc}, using the approximate boundaries of the cloud seen in Figure \ref{fig:g5map} given by the Mopra HOPS \ce{NH3} (3,3) Survey \citep{hops1,hops2,hops3}. 
Two smaller clouds were found to make up G5 within our fields as shown in Figure \ref{fig:g5map}. The first cloud, which we designate G5a, encompasses Field 1 and the east side of Field 2. The fields cover a projection of $\sim$ \SI{22.0}{pc} in Galactic latitude and $\sim$ \SI{24.4}{pc} in Galactic longitude. The second cloud we designate G5b, which takes up the west side of Field 2. Figure \ref{fig:spectra} shows the averaged spectra across the two fields. 

We received 16 Total Power (TP) spectral cubes imaged using the ALMA pipeline calibration and imaging. We checked the cubes over for any flaws in the data cubes\footnote{
An atmospheric feature present in the H(30)$\alpha$ cube resulted in the cube being unusable, as the intensity of the atmospheric feature drowned out any emission from H(30)$\alpha$. 
}.

Next, we checked the rest frequencies of the spectral windows to ensure that they matched those that were targeted. We recorded the approximate velocities of notable spectral features.

\subsection{Continuum Fitting}

We found that Field 1 had poor baseline flatness. We used the CASA task \texttt{imcontsub} to fit a low-order polynomial to line-free channels, then subtracted the continuum model from the data.

For the \ce{CO} isotopologues \ce{^12CO} and \ce{^13CO}, there were too few channels without lines to fit with a continuum model in Field 1. The poor baseline flatness caused dips in the spectrum in different spatial locations of the cubes, especially for \ce{^12CO}, causing the velocity integrated intensity in effected areas to be lowered. 
We masked out negative values, which are a result of baseline-oversubtraction, when creating moment maps (\S \ref{sec:momentmaps}) and for line ratio measurements.
This created an artifact in the \ce{^12CO} data at $(\ell,b) = (+5.413, -0.3725)$.
We did not use Field 1 for any mass measurements, so the poor baselines in the field only affect our figures, not our measurements.

\subsection{Baseline Ripple}
\label{sec:ripple}

We found that Field 2 shows a ripple in the spectral axis, which must be a residual of the baseline removal in the ALMA Single Dish pipeline.
Single dish data often suffers from unstable baselines. Baseline ripples originate from multi-path reflections off of the structure of the telescope from a bright radio source, and can also occur in cables and other pieces of electronics. These reflections cause a standing wave in the optics, which makes a sine wave appear in each spectrum. 
The ALMA data reduction team had done baseline correction before imaging, however we found that Field 2 still showed residual ripples in the spectral cubes). 

In an effort to remove the baseline ripple, the \texttt{numpy.percentile} function was used on the cubes to find the nth percentile of the data, with n varying between values of 1, 5 and 10 depending on the difference in strength between the ripple and the line. 
The nth percentile was determined by examining the output to ensure that no real data was being removed, while still removing the baseline ripple. 
Since the baseline ripple was constant spatially per cube, the nth percentile was then subtracted from each pixel. 
The percentile subtracted from each cube affected by baseline ripple is listed in Table \ref{tbl:persub}. 
The percentile subtraction method shifts the baseline away from zero, so a constant value was subtracted from each cube individually to return the baselines to zero. 
The vertical shift after percentile subtraction is listed in Appendix Table \ref{tbl:persub} for each cube affected by the baseline ripple. 
Removing the baseline ripple revealed dark structures on the resulting position velocity diagrams, constant spatially. 
The process of removing the baseline ripple is shown in Appendix Figure \ref{fig:percentile}.

\subsection{Field Combination}

We mosaicked the image data for the two fields to create one image. 
We combined the two fields by first finding a combined world coordinate system (WCS) and shape containing both fields using the \texttt{reproject} task \texttt{find\_optimal\_celestial\_wcs}. Next, we created a new header of the resulting combined field by editing a copy of the existing header for one of the fields. We then used \texttt{spectral\_cube}'s \texttt{reproject} function to regrid the cubes to the same WCS. We used the masks of the cubes to come up with a weighting grid, so that where the fields overlapped was valued at 2. We ran a loop over each channel of the cubes to combine them by adding each slice together and then dividing by the weighting grid to take the mean of the overlapping points. We then used the resulting array of values and the new header to create a combined cube containing both fields of G5.

\subsection{Additional Molecular Line Detection}

Additional molecular lines were detected in the selected spectral windows. These lines are listed in Table \ref{tab:obs}. For completeness, we identify these lines here. 

In the \ce{H2CO} $J=3_{21}\shortrightarrow2_{20}$ spectral window, the additionally detected line is \ce{OCS} $J=18\shortrightarrow17$. The \ce{OCS} line does not interfere with the \ce{H2CO} line, but only a component of the line is included in the cube. 

In the \ce{HC3N} $J=24\shortrightarrow23$ spectral window, there is no detection of \ce{HC3N}. 
Instead, there are incomplete detections of a component of \ce{CH3OH} $J=4_{22}\shortrightarrow3_{12}$,  \ce{p-H2CO} $J=3_{22}\shortrightarrow2_{21}$, and \ce{H2CO} $3_{03}\shortrightarrow2_{02}$. The \ce{HC3N} cube spectrally overlaps the \ce{H2CO} $3_{03}\shortrightarrow2_{02}$ cube. 

No further analysis was performed on the additional molecular lines. 

\section{Results}
\subsection{Moment Maps}
\label{sec:momentmaps}

\begin{figure*}[htp]
\centering
    \includegraphics[width=0.4\textwidth]{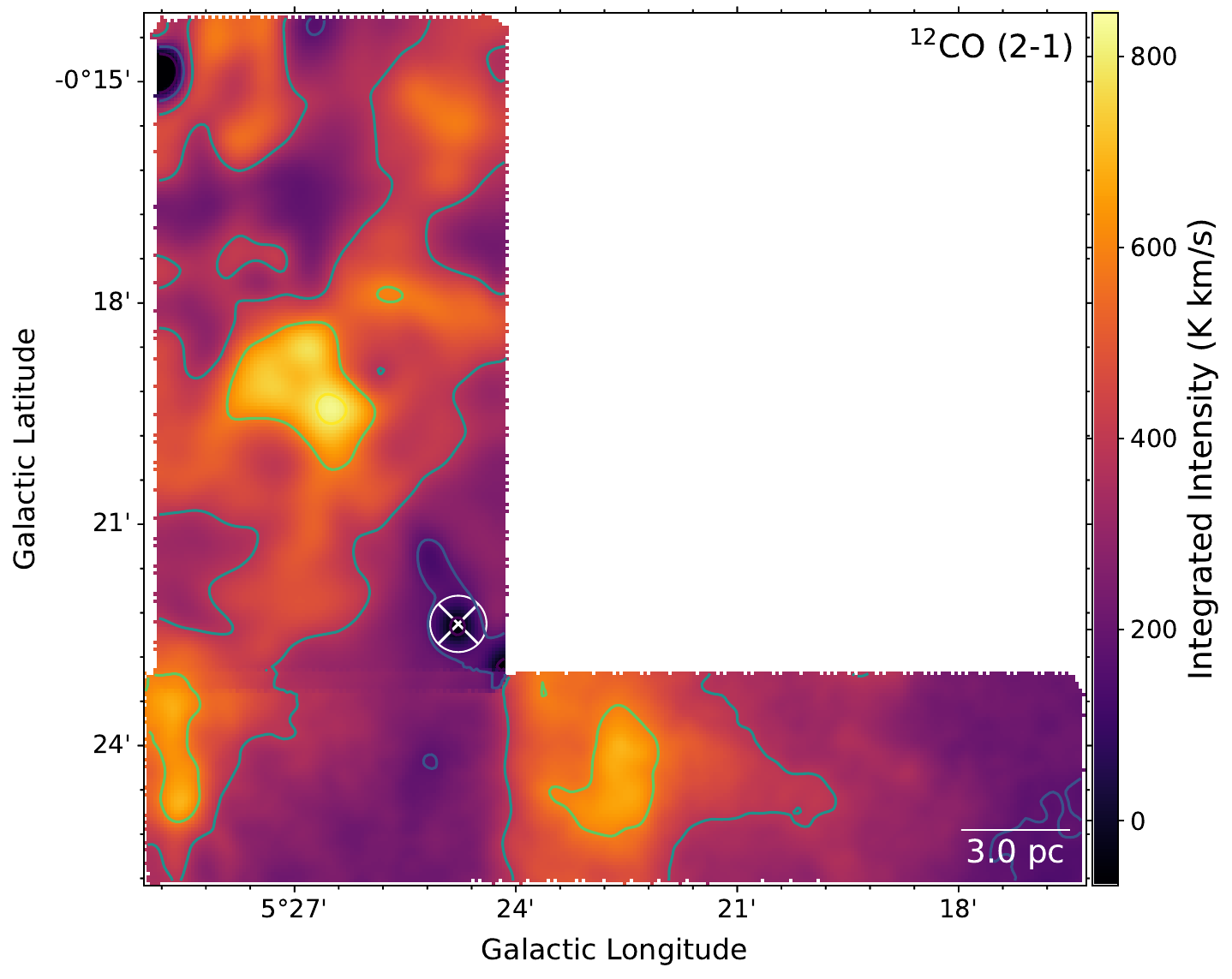}
    \includegraphics[width=0.4\textwidth]{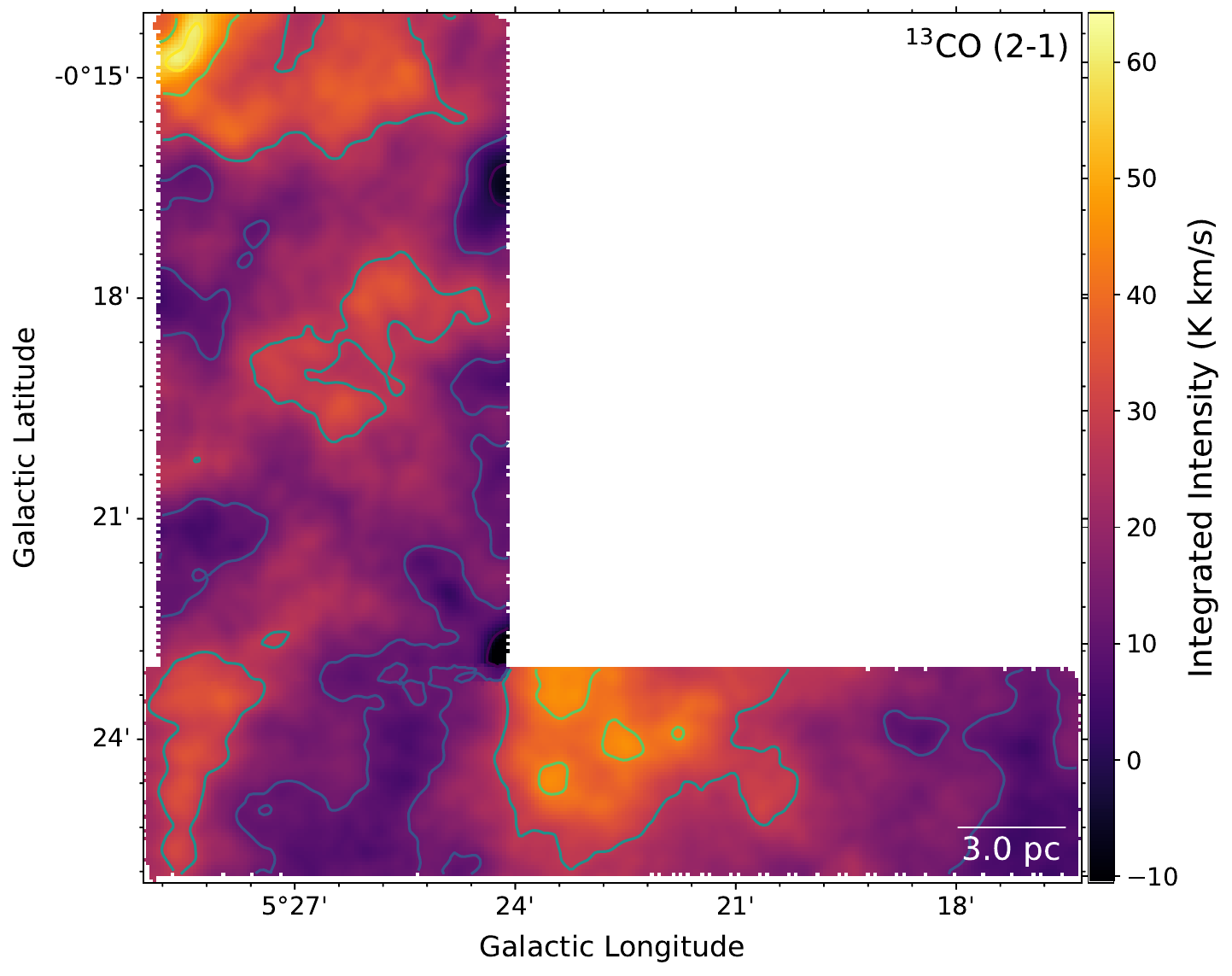}
    \includegraphics[width=0.4\textwidth]{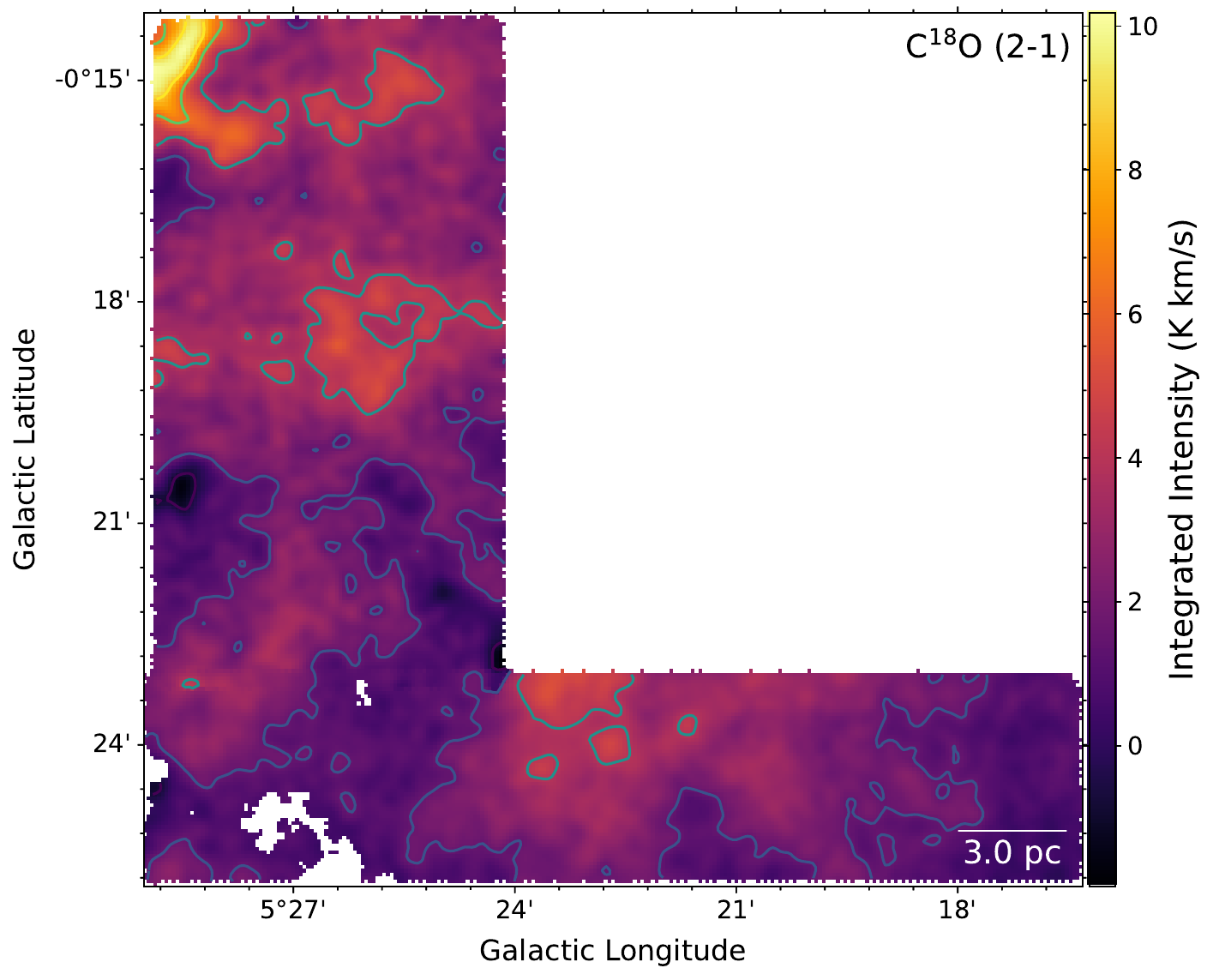}
    \includegraphics[width=0.4\textwidth]{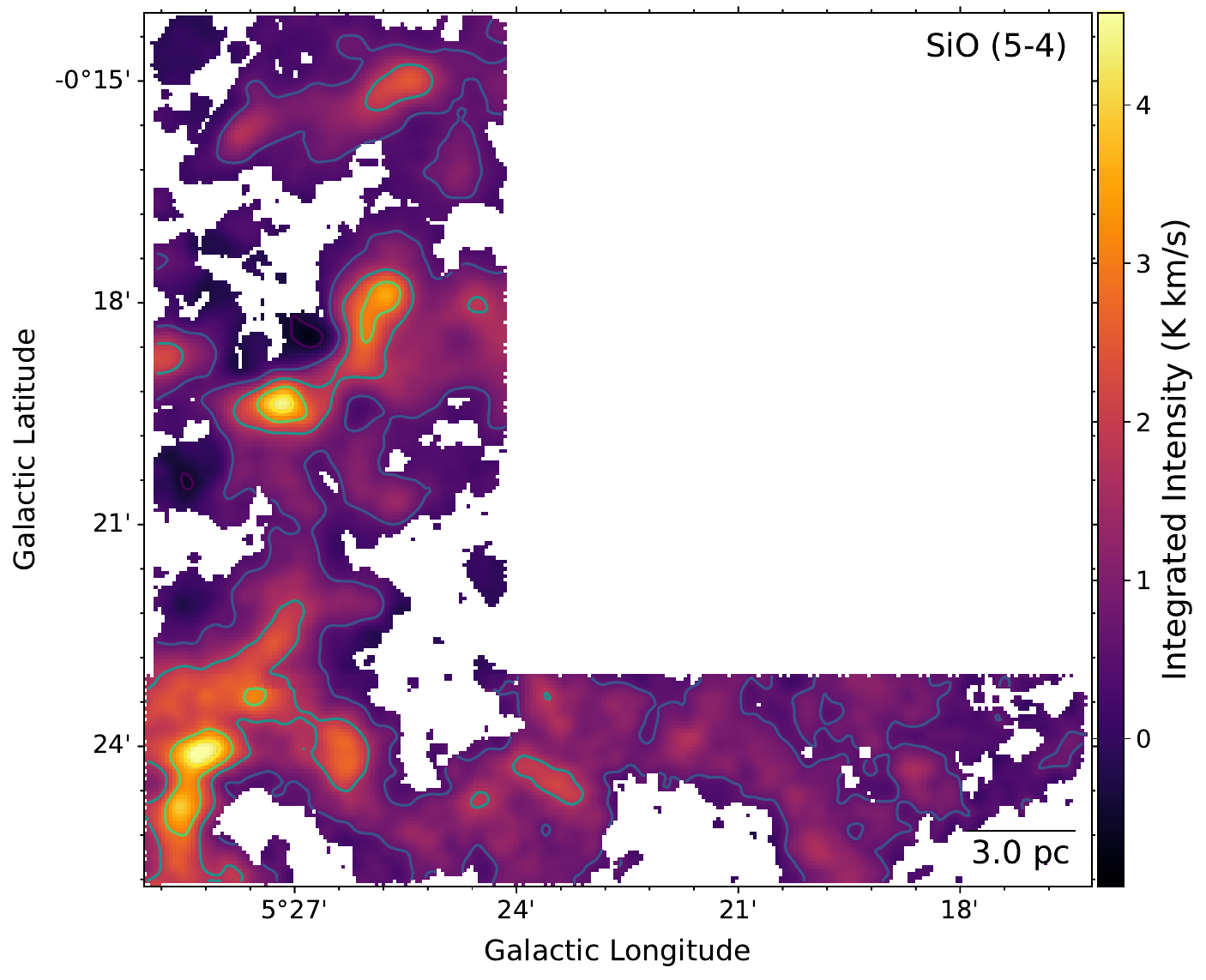}
    \includegraphics[width=0.4\textwidth]{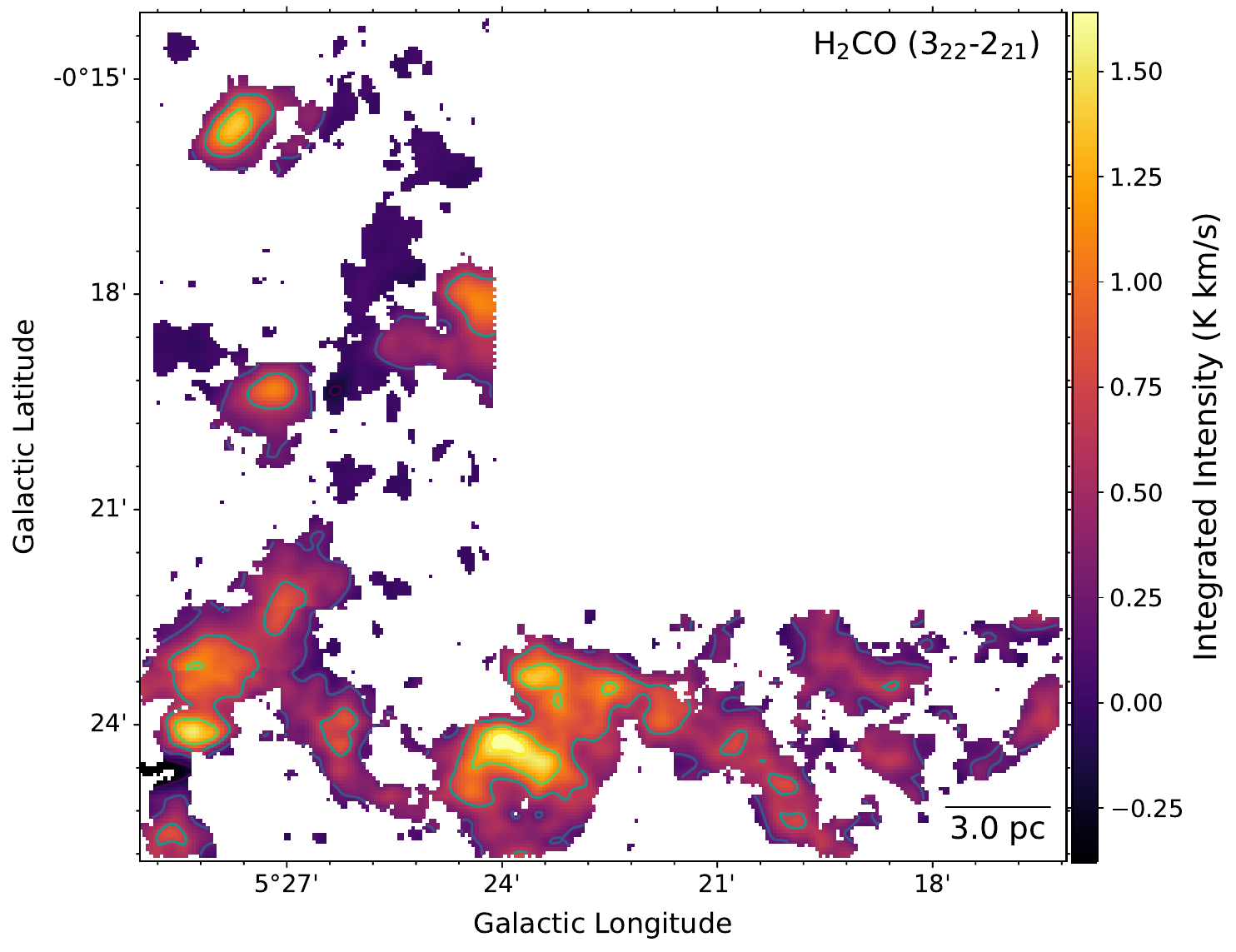}
    \includegraphics[width=0.4\textwidth]{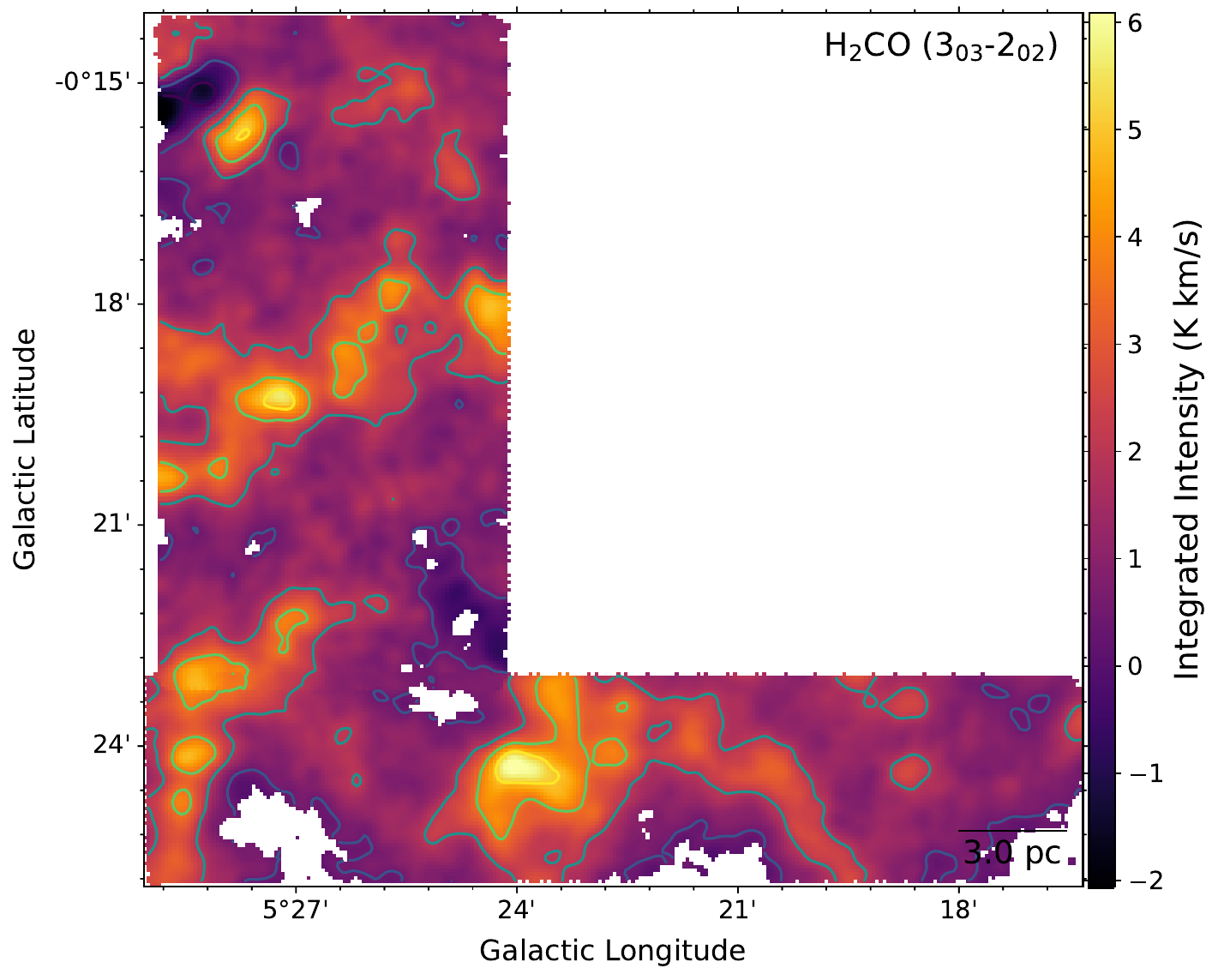}

    \caption{Integrated Intensity maps of the surveyed region of G5. 
    First row: left image is 
    \ce{^12CO} $J=2\shortrightarrow1$, right image is 
    \ce{^13CO} $J=2\shortrightarrow1$. 
    Second row: \ce{C^18O} $J=2\shortrightarrow1$, 
    \ce{SiO} $J=5\shortrightarrow4$.
    Third row: \ce{H2CO} $J=3_{22}\shortrightarrow2_{21}$, 
    \ce{H2CO} $J=3_{03}\shortrightarrow2_{02}$. 
    The contours are the integrated intensity with five contours evenly spaced between the 0.25th and 99.75th percentiles. 
    \label{fig:mom0}}
\end{figure*}

Integrated intensity and velocity field maps were obtained for G5 using methods from the \texttt{spectral-cube} package for \texttt{moment0}  and \texttt{moment1}. 
We first spatially masked the cubes by considering only pixels where the peak was greater than 5 times the noise estimated from the median absolute deviation, then found the integrated intensity. We obtained the second moment and converted it to a FWHM using \texttt{linewidth\_fwhm} from \texttt{spectral-cube} to make an intensity weighted velocity dispersion map, which computes a full-width at half maximum (FWHM) line width map across the spectral axis. 
 
 The moment 0, or integrated intensity, maps of all observed spectral lines are presented in Figure \ref{fig:mom0}. The \ce{CO} isotopologues all share a similar structure with a large feature in the middle of Field 1 stretching across it horizontally. There is also a feature at the bottom of Field 1 that extends into the Galactic east side of Field 2 across the averaged portion of the map smoothly. At the Galactic northeast corner of Field 1 is a very bright region of \ce{^13CO} $J=2\shortrightarrow1$ and \ce{C^18O} $J=2\shortrightarrow1$ that shows up as a feature with absorption in \ce{^12CO} $J=2\shortrightarrow1$. This part of the map overlaps with an HII region, as seen in Figure \ref{fig:g5map} with the 70 $\mu$m Hi-GAL \citep{higal2016} contours. The HII region is not associated with G5a. The velocity of the HII region is \SI{28.3}{\kms} \SI{\pm 0.9}{\kms} with a FWHM of \SI{20.8}{\kms} \SI{\pm 2.0}{\kms} \citep{wink83}, which differs from that of G5a, which has velocities between \SI{50}{\kms} and \SI{100}{\kms} where the HII region overlaps. 

 \begin{figure*}[htp]
     \centering
     \includegraphics[width=.45\textwidth]{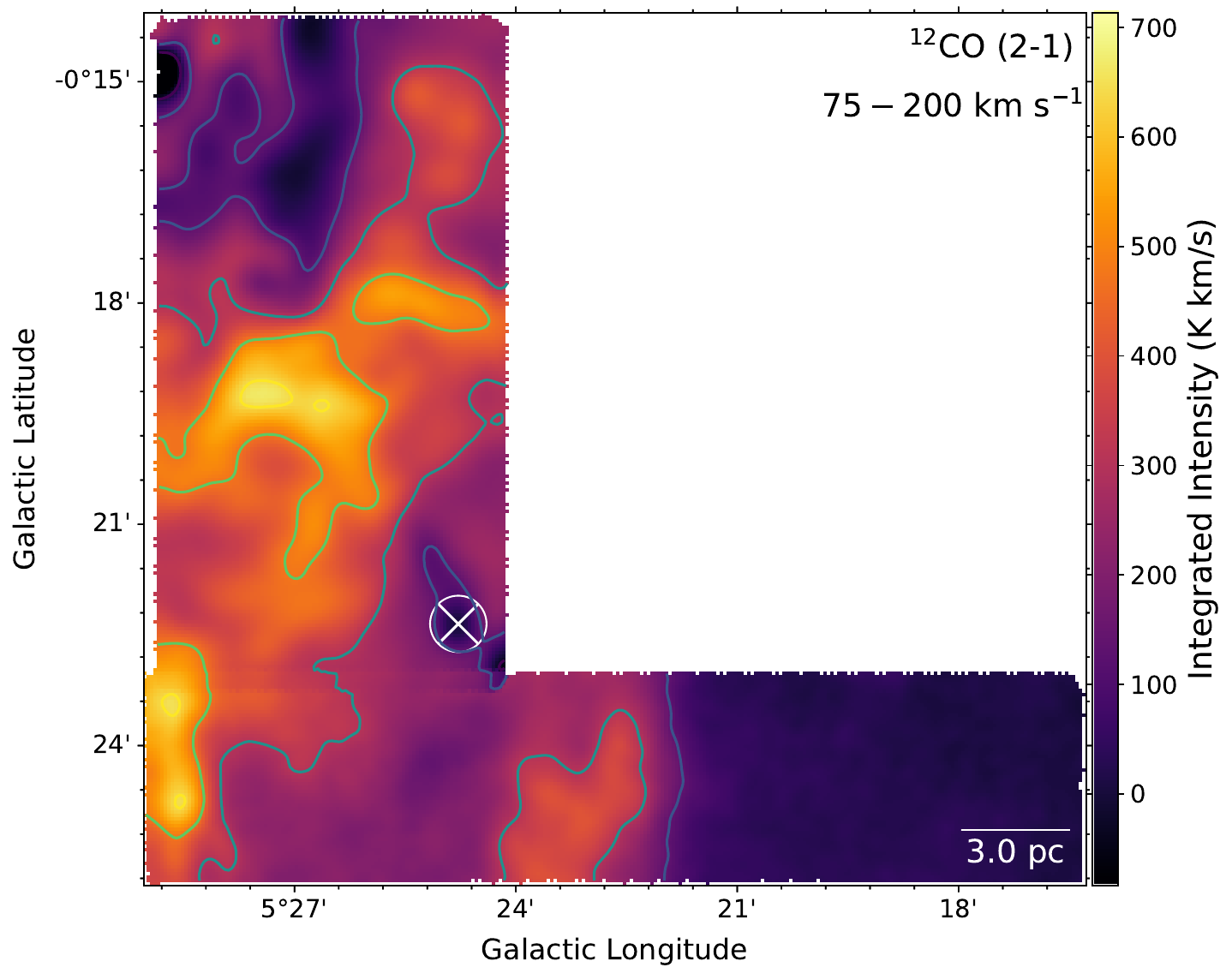}
     \includegraphics[width=.45\textwidth]{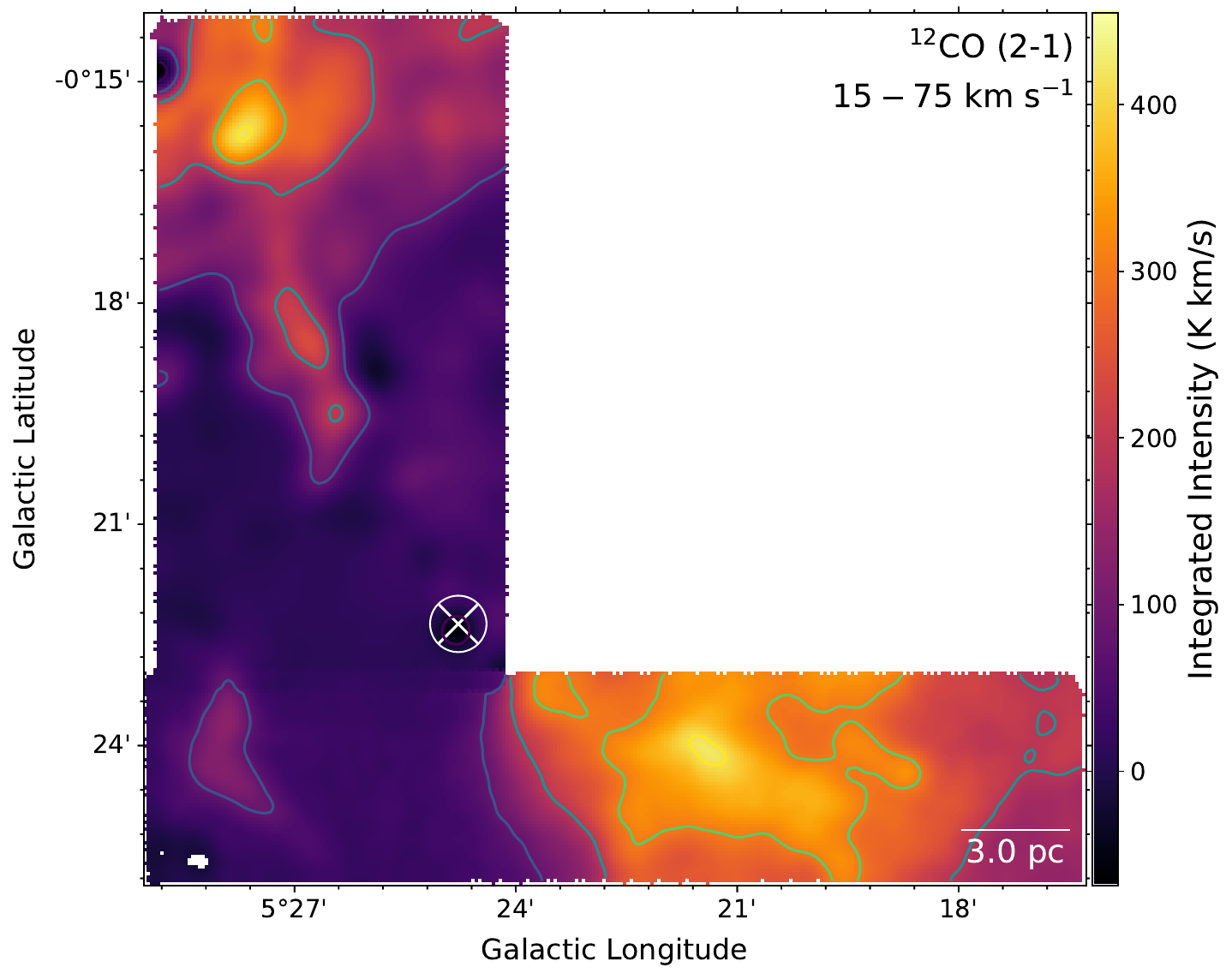}
     \caption{Two integrated intensity maps of G5 in \ce{^12CO} $J=2\shortrightarrow1$ to show the distinct velocity components. 
     Top: Integrated intensity map of G5a, from velocities \SI{75}{\kms} to \SI{200}{\kms}.
     Bottom: Integrated intensity map of G5b, from velocities \SI{15}{\kms} to \SI{75}{\kms}. This map was taken starting from \SI{15}{\kms} to avoid emission from line of sight clouds in the Milky Way's spiral arms.
     }
     \label{fig:separate}
 \end{figure*}
 
 Figure \ref{fig:separate} shows integrated intensity maps separated into chunks from 
 \SI{75}{\kms} to \SI{200}{\kms} for G5a
 and
 \SI{15}{\kms} to \SI{75}{\kms} for G5b.
 The top image of G5a shows a velocity component that runs down the length of Field 1 and down into Field 2. The bottom image of G5b also shows some of the \ce{CO} emission likely associated with the overlapping HII region in Field 1.

\begin{figure*}[ht!]
    \centering
    \includegraphics[width=.45\textwidth]{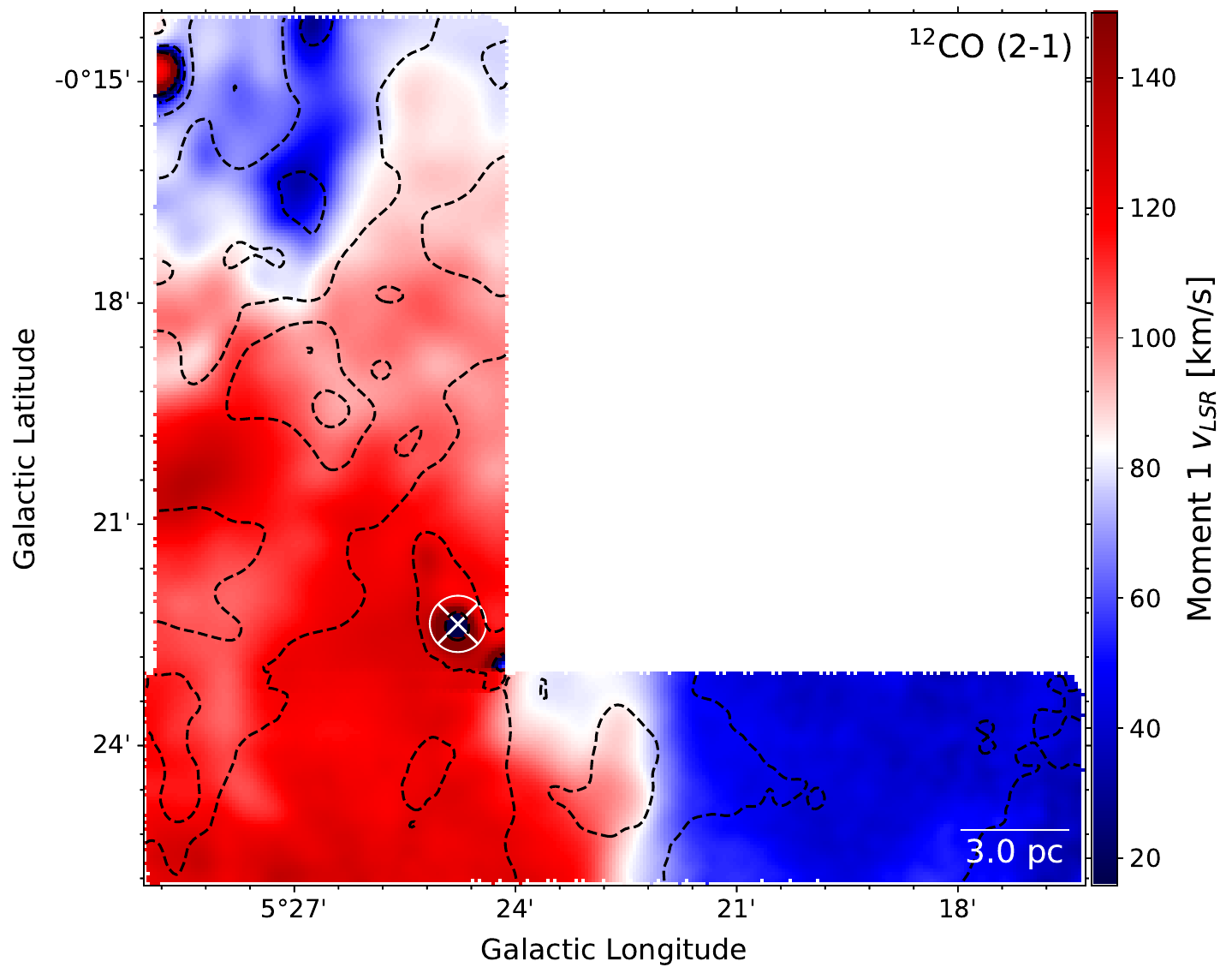}
    \includegraphics[width=.45\textwidth]{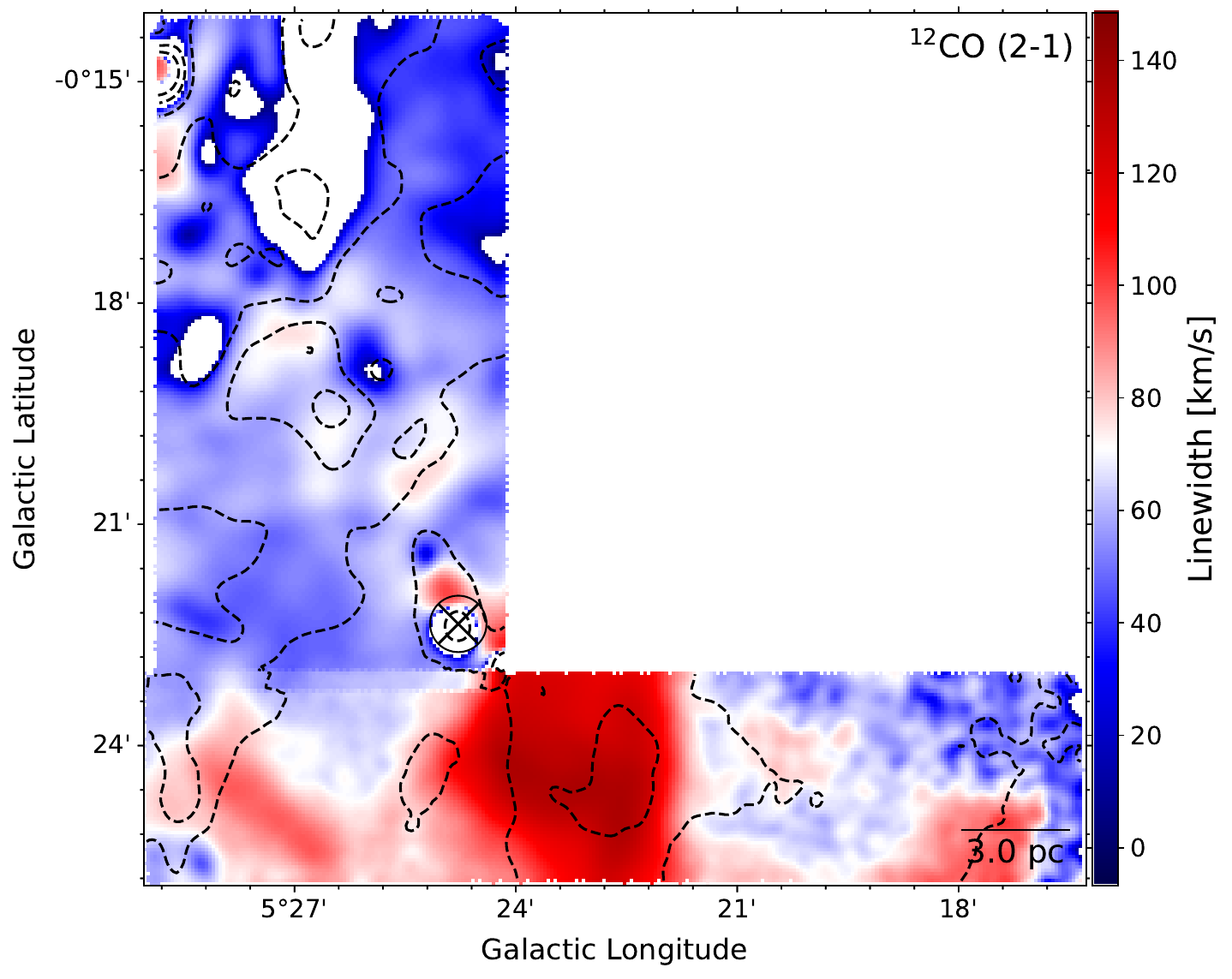}
    \caption{
    (Left) Velocity field map of \ce{^12CO} $J=2\shortrightarrow1$. The red marks the high velocity cloud, and darker blue the low velocity cloud. The lighter blue in the Galactic northeast of the plot is gas associated with the overlapping HII region. Note that the HII region has a different velocity from the nearby G5 gas. 
    (Right) Velocity dispersion map of \ce{^12CO} $J=2\shortrightarrow1$. The elevated dispersion in red marks the interface between the two clouds where they overlap. Note that the elevated dispersion is due to overlapping spectral features, see Figure \ref{fig:all_pvdiagrams}. There is not a very wide spectral feature over all of the area. 
    Black dashed line contours are \ce{^12CO} $J=2\shortrightarrow1$ integrated intensity with five contours evenly spaced between the 0.25th and 99.75th percentiles.
    \label{fig:mom12}}
\end{figure*}

The left image in Figure \ref{fig:mom12} is a velocity field map of G5. The figure clearly shows that G5 contains two major velocity components. The red G5a in the Galactic east and the blue G5b in the Galactic west are separated by a white transition between the velocity components, which approximately lines up with a peak in \ce{^12CO} $J=2\shortrightarrow1$ as shown by the contours. 

The light blue region in the Galactic northeast of Figure \ref{fig:mom12} is gas associated with the HII region along the same line of sight as G5. 

The right image in Figure \ref{fig:mom12} is a velocity dispersion map of G5. The blue shows the FWHM where the clouds are not overlapping or interacting. The red marks a heightened region of velocity dispersion due to the overlap and interaction between the two clouds. The apparently large velocity dispersion in the center of the image, caused by two separate velocity components, of up to $\sim$ \SI{150}{\kms} should not be confused with internal dispersion of the gas in the clouds. 
The large velocity dispersion is reflecting the velocity gap between the two spatially overlapping detections of \ce{^12CO} $J=2\shortrightarrow1$, not the FWHM of the molecular lines.
The typical line widths in G5 are on the order of \SI{30}{\kms} to \SI{50}{\kms}.

\subsection{Position-Velocity Diagram}
\label{sec:pvdiagrams}

\begin{figure*}[htp]
    \centering
    \includegraphics[width=1\textwidth]{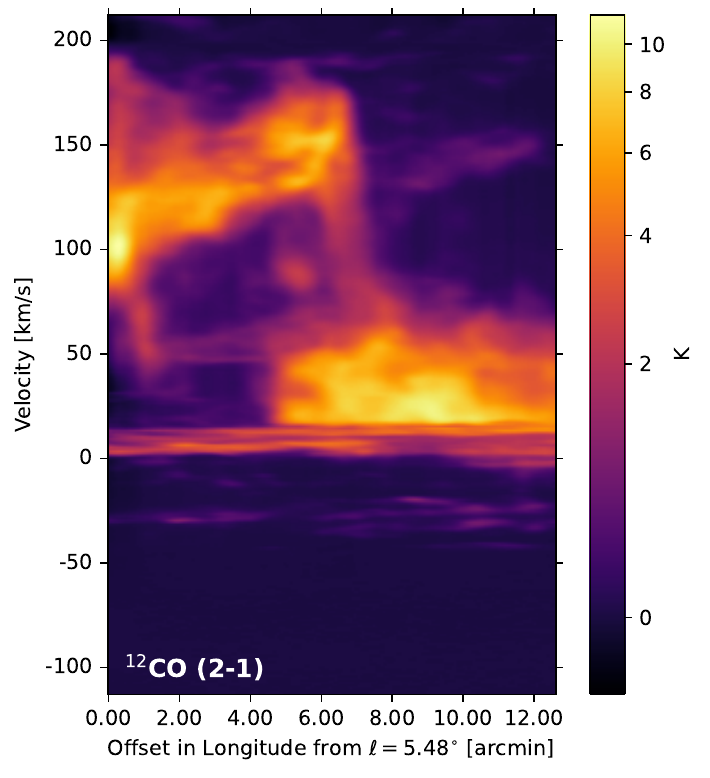}
    \caption{Position-Velocity diagram of Field 2 in \ce{^12CO} $J=2\shortrightarrow1$, averaged over Galactic latitude and taken horizontally across the field with a width of \SI{2}{\arcmin}. Features are labelled in Appendix Figure \ref{fig:labelled_pv}.
    \label{fig:pvdiagram}}
    
\end{figure*}

\begin{figure*}[htp]
\centering
    \vspace{-5pt}  
    \includegraphics[width=0.39\textwidth]{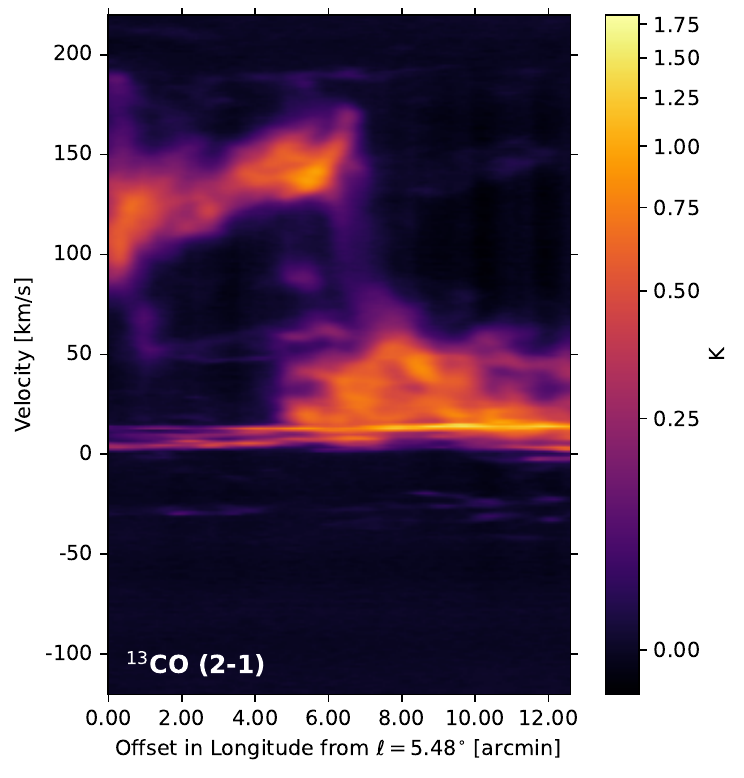}
    \hspace{1cm}
    \includegraphics[width=0.39\textwidth]{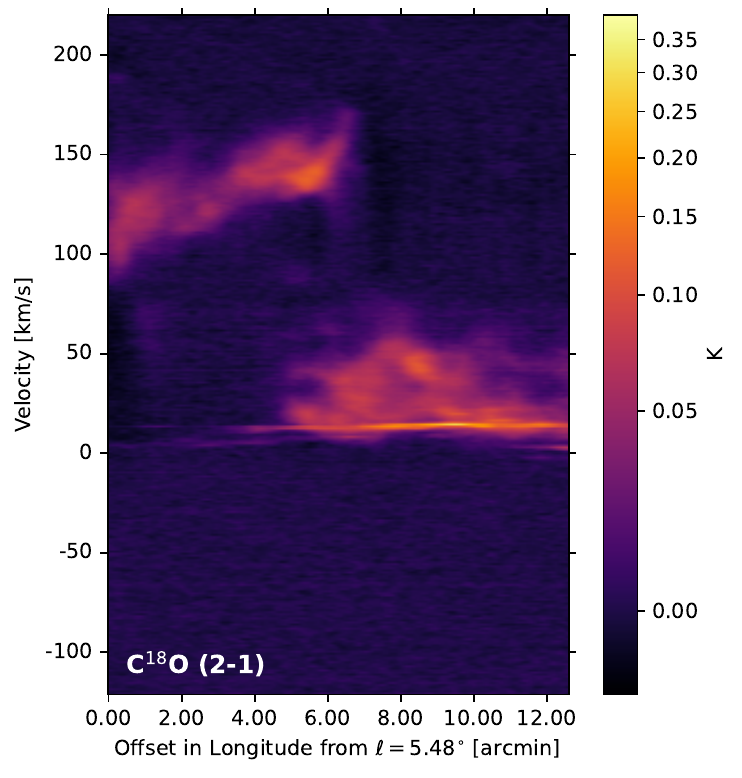}
    \includegraphics[width=0.39\textwidth]{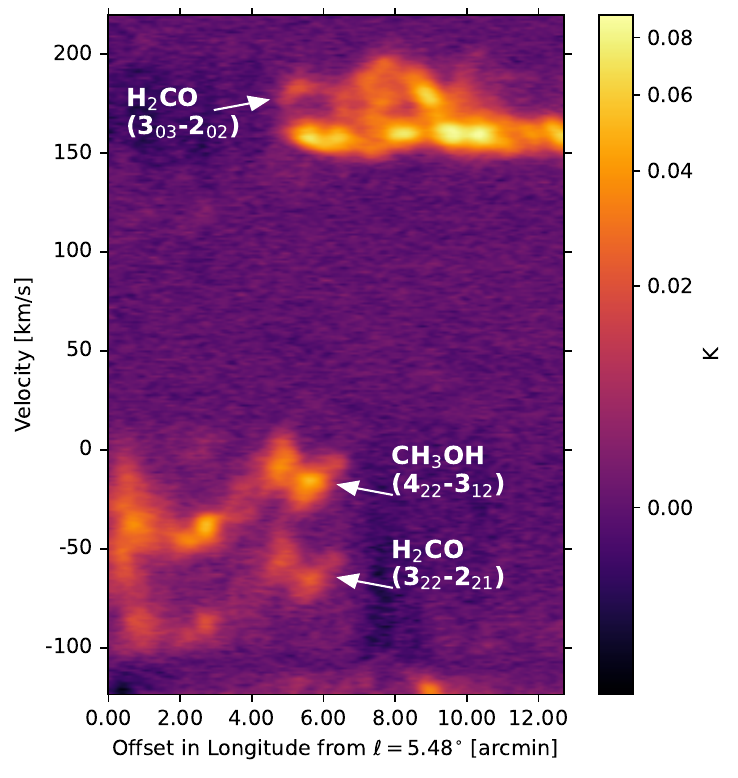}
    \hspace{0.7cm}
    \includegraphics[width=0.41\textwidth]{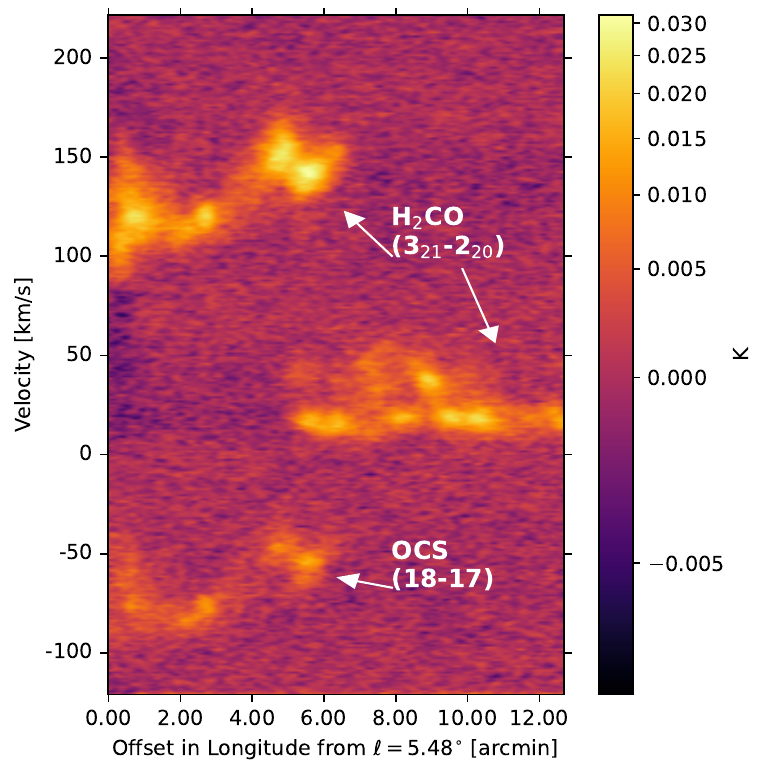}
    \includegraphics[width=0.39\textwidth]{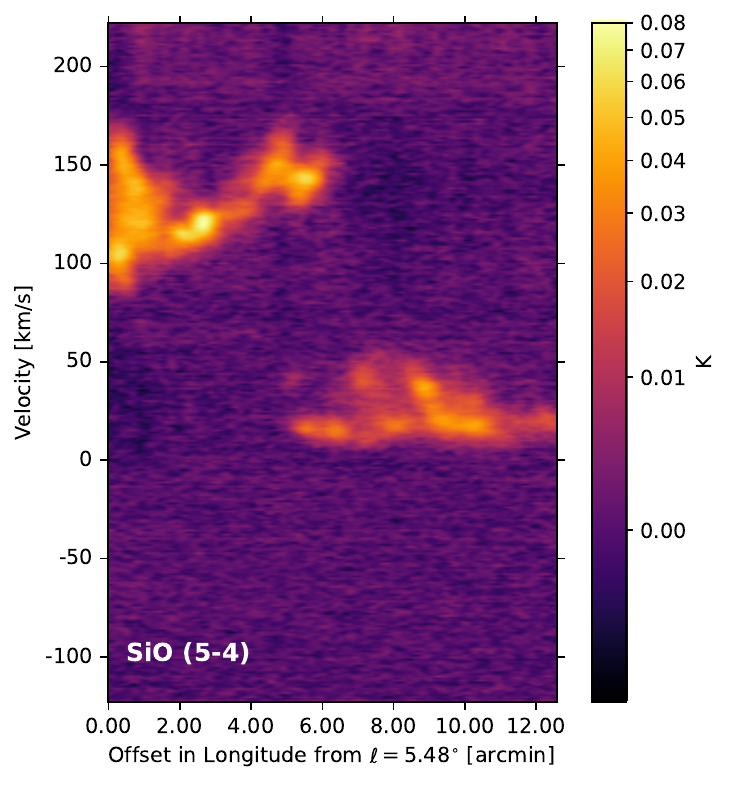}
    \hspace{1cm}
    \includegraphics[width=0.39\textwidth]{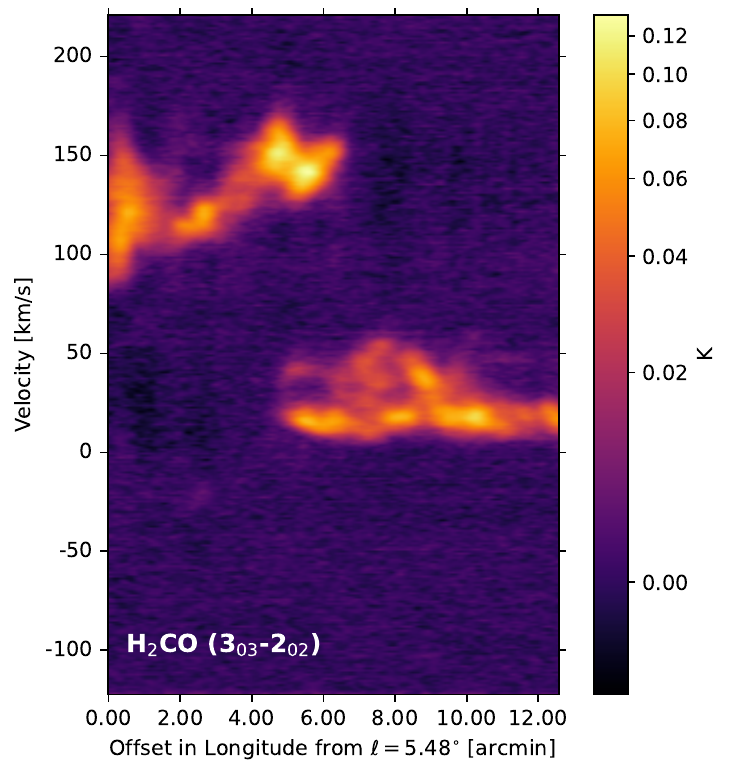}

    \caption{Position-Velocity Diagrams of Field 2 of G5, scaled by asinh. 
    First row: first image is 
    \ce{^13CO} $J=2\shortrightarrow1$ second image is \ce{C^18O} $J=2\shortrightarrow1$. 
    Second row: 
    \ce{HC3N} $J=24\shortrightarrow23$,
    \ce{H2CO} $J=3_{21}\shortrightarrow2_{20}$. 
    Third row: \ce{SiO} $J=5\shortrightarrow4$, \ce{H2CO} $J=3_{03}\shortrightarrow2_{02}$.
    The PV Diagram of \ce{HC3N} has no detection of \ce{HC3N}, but it is contaminated with emission from 
    \ce{H2CO} $J=3_{03}\shortrightarrow2_{02}$ ($\SI{150}{\kms}$ to $\SI{200}{\kms}$), 
    \ce{CH3OH} $J=4_{22}\shortrightarrow3_{12}$ ($\SI{-50}{\kms}$ to $\SI{0}{\kms}$) and 
    \ce{H2CO} $J=3_{22}\shortrightarrow2_{21}$ ($\SI{-100}{\kms}$ to $\SI{-50}{\kms}$). 
    The PV Diagram of \ce{H2CO} \ce{H2CO} $J=3_{21}\shortrightarrow2_{20}$ is contaminated with emission from \ce{OCS} $J=18\shortrightarrow17$ at velocities between $\SI{-100}{\kms}$ and $\SI{-50}{\kms}$. 
    \label{fig:all_pvdiagrams}
    }
\end{figure*}

We created a position-velocity (PV) diagram, Figure \ref{fig:pvdiagram}, by selecting a range of data horizontally in Galactic Longitude across Field 2 for the \ce{^12CO} $2\shortrightarrow1$ cube, with a width of \SI{2}{\arcmin}. The offset of Figure \ref{fig:pvdiagram} is relative to the left of Field 2 at $\ell=$\SI{5.48}{\degree}, so \SI{0}{\arcmin} is at a higher Galactic longitude. PV diagrams for the other observed spectral windows are shown in Figure \ref{sec:pvdiagrams}.

We identify several features in the PV diagram in Figure \ref{fig:pvdiagram}. We first find G5a on the left side of the field at  $\sim$\SI{150}{\kms}, stretching from a position offset of $\sim$\SI{0}{\arcmin} to $\sim$\SI{6}{\arcmin}. A second cloud with a wide velocity dispersion on the right side of the field is identified as G5b, which is at $\sim$\SI{50}{\kms} but stretches to $\sim$\SI{15}{\kms}, from a \SI{5}{\arcmin} to \SI{12}{\arcmin} offset. Stretching between the two clouds in the velocity domain at an offset of $\sim$ \SI{7}{\arcmin} is a velocity bridge. 

A velocity bridge is a feature in a PV diagram which is wide in velocity space but relatively narrow in position space, and connects the two features at $\sim$\SI{150}{\kms} and at $\sim$\SI{50}{\kms}, spatially the velocity bridge is where the two clouds overlap \citep{HaworthDec15, HaworthJun15}.
The \ce{^12CO} PV diagram in Figure \ref{fig:pvdiagram} has a vertical velocity bridge connecting the two clouds. 
We discuss the details and implications of the velocity bridge feature in Section \ref{sec:velocitybridge}.


The clump at offset \SI{5}{\arcmin} and at the velocity \SI{80}{\kms} could be associated with a secondary bar lane feature identified in \citet{sormani19} and \citet{Liszt2006}, or it is somehow associated with the velocity bridge.

We find another extended velocity feature of similar spectral width to the velocity bridge on the left side of the field at approximately $(\ell,b) = (5.47, -0.41)$\SI{}{\degree}, which seems connected to G5a, but does not seem to directly link it to the G5b. The feature has an unusually wide velocity from $\sim$\SI{30}{\kms} to $\sim$\SI{100}{\kms} where it intersects with G5a at around a \SI{0.5}{\arcmin} offset. While this feature does not seem to directly intersect with G5b, there is a cloud feature at $\sim$\SI{50}{\kms} that it may be interacting with. 
This `spur' off the main body of G5a has not been conclusively identified. 
One potential explanation is that the spur is due to a secondary bar lane feature identified in \citet{sormani19}, or it is evidence of a different cloud interaction at G5 similar to the velocity bridge.

There are also several sources of emission with very narrow velocity dispersion at $\sim$\SI{0}{\kms} to \SI{15}{\kms} and at \SI{35}{\kms}, which we believe are foreground molecular clouds in the Milky Way's disk.

\clearpage
\subsection{Kinetic Temperature} 
\label{sec:gastemperature}

\begin{figure}[htp]
    \centering
    \includegraphics[width=.45\textwidth]{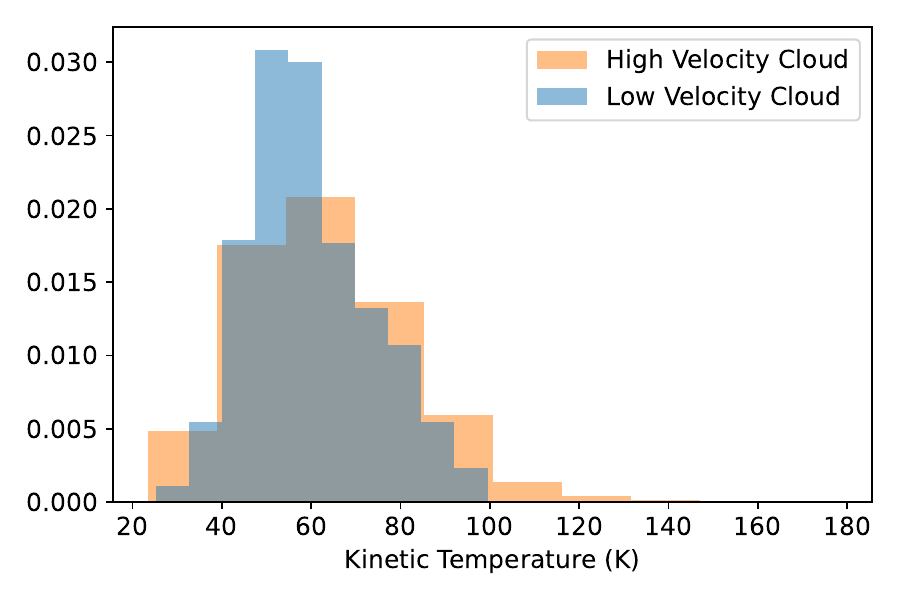}
    \caption{Histogram of the gas temperatures of G5a, the high velocity cloud in orange, and G5b, the low velocity cloud in blue. 
    The average temperatures of the two clouds are \SI{63}{K}$\pm$\SI{19}{K} for G5a and \SI{60}{K}$\pm$\SI{14}{K} for G5b. 
    This figure shows how much of each cloud is in each temperature bin, separated between G5a and G5b. 
    A KS test between the two clouds resulted in a p-value $<<$ 1, meaning that the temperatures of G5a and G5b do not come from the same distribution.}
    \label{fig:temphist}
\end{figure}

We determine the temperature of G5 using a line ratio between \ce{H2CO} $J=3_{22}\shortrightarrow2_{21}$ and \ce{H2CO} $J=3_{03}\shortrightarrow2_{02}$ integrated intensity maps. We find the temperatures for G5a and G5b in Field 2 separately by making integrated intensity maps which covered only the velocity ranges of G5a and G5b. 
The two clouds are spatially superimposed in the center of Field 2, making the integrated intensity maps of the whole velocity ranges, in Figure \ref{fig:mom0}, possibly result in erroneous temperature measurements, as two clouds are measured at once. 
There is no detection in \ce{H2CO} of the velocity bridge feature identified in \ce{CO}, so we do not attempt to measure its temperature.

We first separate the cubes of \ce{H2CO} $3_{0,3}\shortrightarrow2_{0,2}$ into subcubes of \SI{75}{\kms} to \SI{225}{\kms} for G5a, and \SI{0}{\kms} to \SI{75}{\kms} for G5b. 
We mask the cubes by considering only pixels along the spectra with a peak signal to noise ratio above 5. 
We find the line ratio between the masked integrated intensity maps of \ce{H2CO} $J=3_{03}\shortrightarrow2_{02}$ and \ce{H2CO} $J=3_{22}\shortrightarrow2_{21}$.

We then use Equation \ref{eq:temp},
\begin{equation} 
    \label{eq:temp}
    T_\mathrm{G} = 590 \times R_{\ce{H2CO}}^2 + 2.88 \times R_{\ce{H2CO}} + 23.4
\end{equation}
a second degree polynomial fit based on a \texttt{RADEX} \citep{radex} model relationship between the line ratio $R_{\ce{H2CO}} = \frac{\int I_{\nu}(3_{21}\shortrightarrow2_{20})d\nu} 
{\int I_{\nu}(3_{03}\shortrightarrow2_{02})d\nu}$ 
and the gas temperature $T_\mathrm{G}$  \citep{ginsburg16}. The model assumes that the volume density of the cloud is \SI{e4}{cm^{-3}}, while being relatively insensitive to the exact volume density.
It also assumes the abundance $X_{\ce{H2CO}}=\SI{1.2e-9}{}$ and an assumed fixed line gradient of \SI{5}{km s^{-1} pc^{-1}}. We solve for the gas temperature of the clouds by putting the line ratio into the equation.

After producing temperature maps of the clouds, we find that there is no significant spatial correlation with kinetic temperature, but we still measure the temperatures of G5a and G5b separately.
We make a histogram of the temperature values found in G5a and G5b, shown in Figure \ref{fig:temphist}.
We find the temperatures of G5a and G5b separately, but the temperatures of the two clouds are relatively similar.
The average temperatures are \SI{63}{K}$\pm$\SI{19}{K} for G5a and \SI{60}{K}$\pm$\SI{14}{K} for G5b. 
The temperatures of G5 are warm compared to the temperatures of non-star forming molecular clouds found in the Galactic disk, measured with \ce{NH3} (1,1) and (2,2), which have temperatures closer to \SI{10}{K} to \SI{20}{K} \citep{Friesen2017}.

\subsection{Shocks}
\label{sec:shocks}

\begin{figure}[tp]
    \centering
    \includegraphics[width=.45\textwidth]{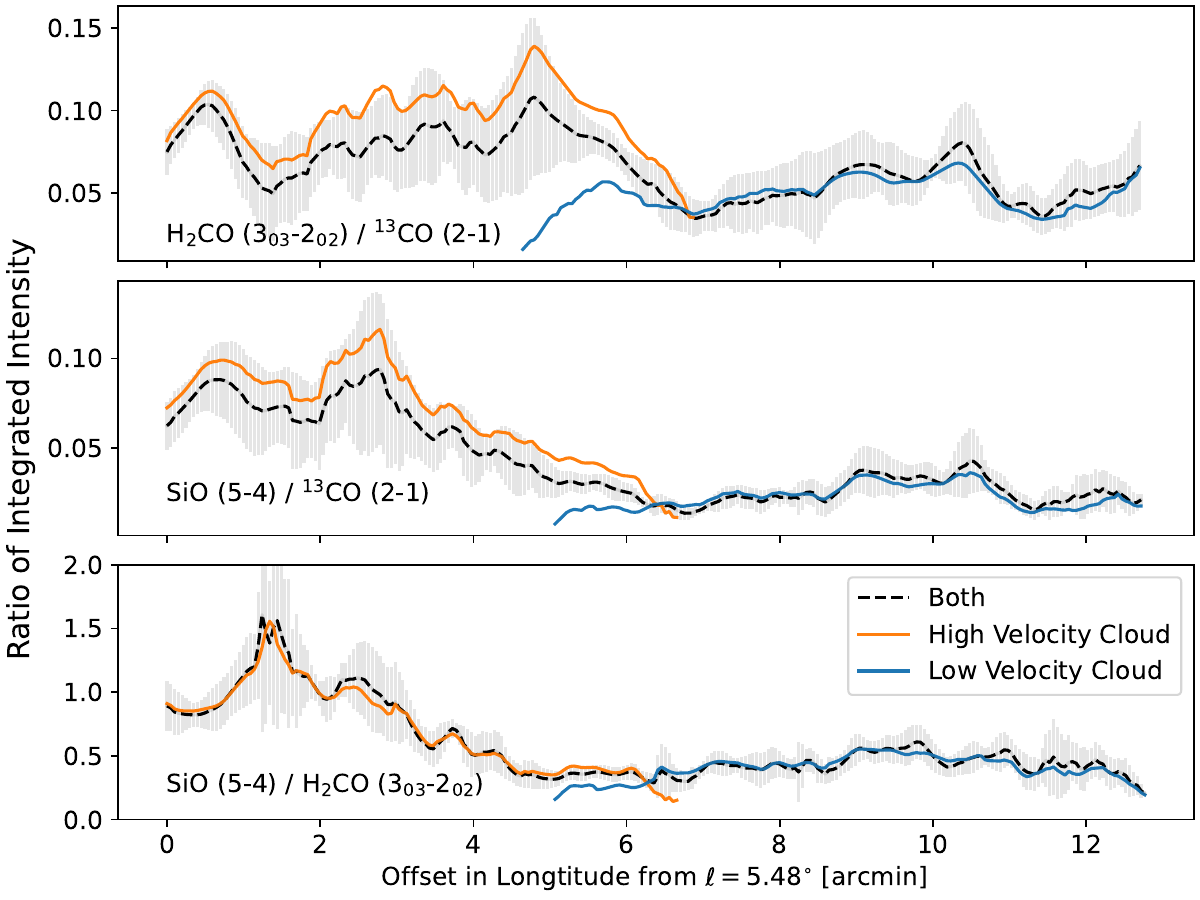}
    \caption{
    Ratios of integrated intensity maps were taken of Field 2 and then averaged over Galactic Latitude to show how the values change horizontally across the field in Galactic Longitude. Black shows the ratios measured by taking the ratio of the integrated intensities over the whole cubes. The grey error bars show the standard deviation over the averaged area. Orange shows the ratios when the integrated intensities were limited to velocities associated with G5a, the higher velocity cloud. Blue shows the ratios when the integrated intensities were limited to velocities associated with G5b, the lower velocity cloud.
    }
    \label{fig:chem}
\end{figure}

\begin{figure}
    \centering
    \includegraphics[width=.45\textwidth]{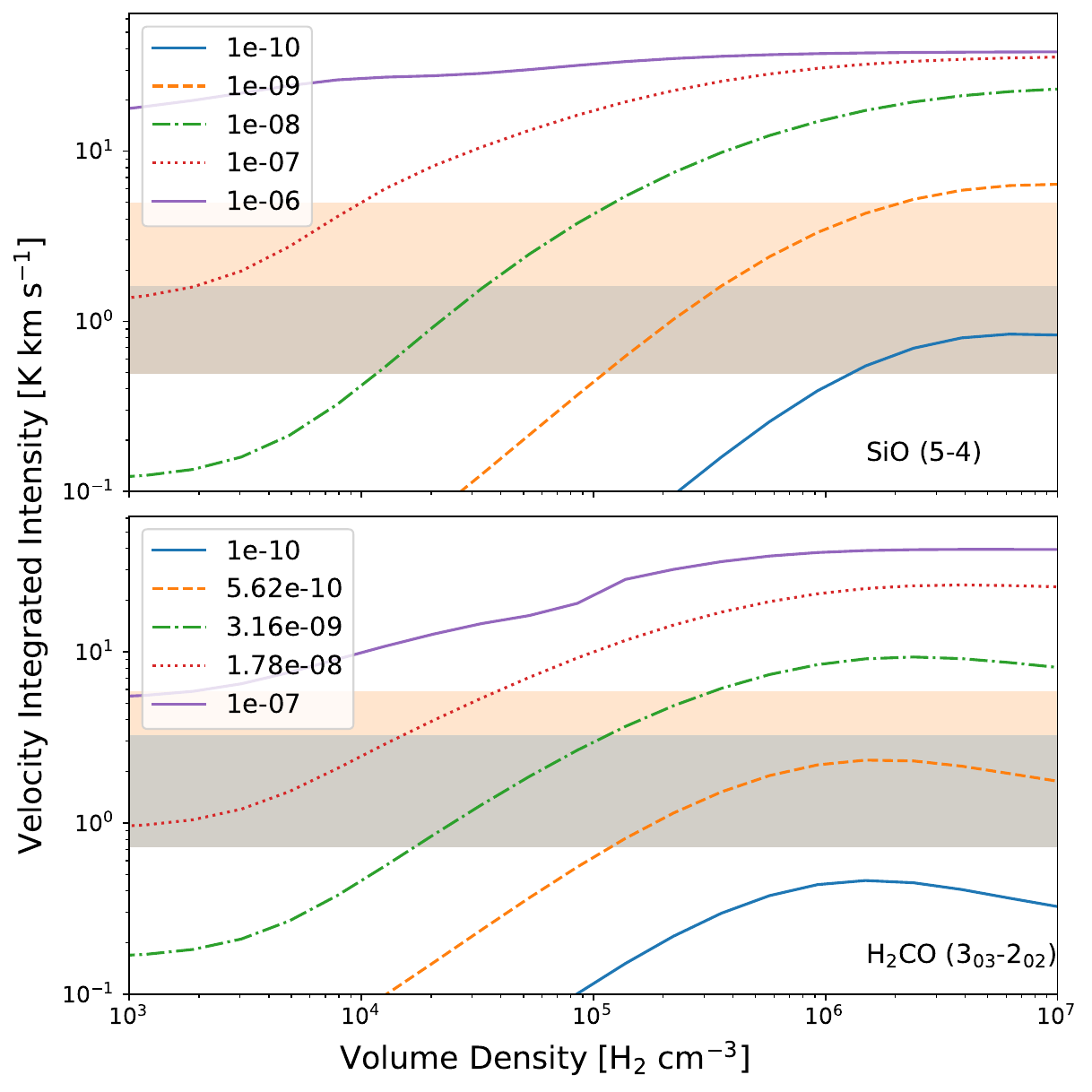}
    \caption{Plot of the simulated integrated intensities of 
    \ce{SiO} $J=5\shortrightarrow4$ (top) and \ce{H2CO} $J=3_{03}\shortrightarrow2_{02}$ (bottom) for different volume densities and abundances at a gas temperature of \SI{60}{K}. 
    The orange span covers the measured integrated intensities for the higher velocity cloud G5a, and the blue span covers the same for the lower velocity cloud G5b. The upper limits of the colored spans are the maximum detected integrated intensities, and the lower limits are the detection threshold.}
    \label{fig:despoticsims}
\end{figure}

To examine the shock properties of G5, we looked at \ce{SiO} $J=5\shortrightarrow4$. \ce{SiO} is a known shock tracer, as shocks are thought to release silicon from cold dust grains into the gas phase, where it chemically interacts with oxygen to produce \ce{SiO}. \citet{schilke97} shows that the abundance of \ce{SiO} increases in strongly shocked regions to \SI{e-7}{} and \SI{e-6}{} compared to the ambient abundance of \SI{e-11}{}. 

In Figure \ref{fig:mom0}, we measure the integrated intensity of \ce{SiO} $J=5\shortrightarrow4$, with a maximum value of \SI{4.8}{K km s^{-1}}.

We compare the line integrated intensity ratios between \ce{^13CO} $J=2\shortrightarrow1$, \ce{H2CO} $J=3_{03}\shortrightarrow2_{02}$, and \ce{SiO} $J=5\shortrightarrow4$. 
We first spatially masked the Field 2 cubes by considering only pixels where the peak was greater than 5 times the noise estimated from the median absolute deviation, then found the integrated intensity. 
We computed the line integrated intensity ratios and then averaged in Galactic Latitude to show the interesting variations in Field 2 between G5a and G5b. 
We plotted the line intensity ratios with errors bars showing the standard deviation of the values averaged in Figure \ref{fig:chem}.

The line intensity ratios with respect to \ce{^13CO} $J=2\shortrightarrow1$ make up the top two panels of Figure \ref{fig:chem}. These ratios give a measure of how much of that molecule is present relative to the amount of gas present.

We compare \ce{SiO} $J=5\shortrightarrow4$ and \ce{H2CO} $J=3_{03}\shortrightarrow2_{02}$ in the third panel of Figure \ref{fig:chem}. 
The critical densities of the two lines are similar, as the critical density of 
\ce{SiO} $J=5\shortrightarrow4$ is \SI{2.52e6}{cm^{-3}} in gas that is \SI{60}{K}, and for 
\ce{H2CO} $J=3_{03}\shortrightarrow2_{02}$ is \SI{3.10e6}{cm^{-3}} \citep{lambda2005}\footnote{Using Leiden Lambda data for \ce{SiO}-\ce{H2} and \ce{pH2CO}-\ce{H2} accessed January 2023.}.
The similarity between their critical densities means that their line intensity ratio is not dependent on density. 
We expect \ce{H2CO} and \ce{SiO} to be optically thin.
While \ce{H2CO} is not insensitive to shocks, \ce{SiO} is expected to be far more sensitive \citep{Bachiller1997}. 
Where the ratio between the two is high, the \ce{SiO} abundance is expected to be higher and enhanced by shocks.

To further compare \ce{SiO} $J=5\shortrightarrow4$ and \ce{H2CO} $J=3_{03}\shortrightarrow2_{02}$, we simulate the lines in non-LTE with \texttt{DESPOTIC} \citep{despotic2014}. We do not detect \ce{HC3N} $24\shortrightarrow23$ anywhere in G5, meaning the gas does not reach densities high enough to excite the line. The approximate upper limit on the volume density of G5 is the critical density of the \ce{HC3N} line at the gas temperature \citep{mills2018}. Since we observe the gas at a temperature of \SI{60}{K} in Section \ref{sec:gastemperature}, the critical density and upper limit on the volume density is \SI{1.75e7}{cm^{-3}} \citep{lambda2005}.
The simulation assumes an \ce{H2} column density of \SI{e22}{cm^{-2}}, gas temperature of \SI{60}{K}, and that the non-thermal velocity dispersion is larger than the sound speed of the gas.

The results of the \texttt{DESPOTIC} simulation are show in Figure \ref{fig:despoticsims}. The plots show how the integrated intensity of the lines change over different volume densities for different abundances of the molecules.

The ratios found in the higher velocity cloud, G5a, are larger than those found in G5b. In some parts of G5a, the \ce{SiO} $J=5\shortrightarrow4$ integrated intensity is even higher than that of \ce{H2CO} $J=3_{03}\shortrightarrow2_{02}$. 
There are two interpretations:
\begin{enumerate}
    \item G5a has a higher excitation due to increased volume density.
    \item G5a has a higher abundance of SiO.
\end{enumerate}
If just the volume density increased from G5b to G5a, we would expect \ce{H2CO} $J=3_{03}\shortrightarrow2_{02}$ to increase just as much as \ce{SiO} $J=5\shortrightarrow4$, but \ce{SiO} increases more. We expect G5a to have a higher abundance of \ce{SiO} than G5b.

The top panel of Figure \ref{fig:despoticsims}, showing the non-LTE simulation for \ce{SiO} $J=5\shortrightarrow4$, shows that for volume densities lower than \SI{e-4}{cm^{-3}}, the line is not excited in the gas. This could explain the absence of \ce{SiO} $J=5\shortrightarrow4$ in the velocity bridge, which is only visible in the PV Diagrams of the \ce{CO} isotopologues. 
If the gas in the velocity bridge is below the required volume density, then \ce{SiO} $J=5\shortrightarrow4$ will not be detected, even at high abundance. The lack of \ce{SiO} $J=5\shortrightarrow4$ in the velocity bridge means that it is not very dense gas. 
Viewing G5 in transitions of \ce{SiO} with a lower $J$ value and critical density could reveal the abundance of \ce{SiO} in the velocity bridge, but the line might not be detectable because of the lowest volume densities.

While we did not detect \ce{SiO} $J=5\shortrightarrow4$ in the velocity bridge, we did detect it in G5a and G5b. 
The higher volume density of the interiors of the clouds allows the line to be excited enough for us to observe it. The top panel of Figure \ref{fig:despoticsims} shows that the abundance of \ce{SiO} can be expected to be within slightly lower than \SI{e-10}{} to over \SI{e-8}{}.


\subsection{Mass Estimation}
    \label{sec:massest}

\begin{table*}
\caption{Mass Estimate of Field 2}
\centering
\begin{tabular}{cccccc}
\hline
Method & Mass Estimate & Column Density (H$_2$) & N(H$_2$)/$\int$ I$_{u}$ ($^{12}CO$) d$u$ & CMZ Inflow & Assumed Ratio \\
 & $\mathrm{M_{\odot}}$ x 10$^5$ & $\mathrm{cm^{-2}}$ x 10$^{22}$ & $\mathrm{cm^{-2}~(K~km~s^{-1})^{-1}}$ x 10$^{20}$ & $\mathrm{M_{\odot}\,yr^{-1}}$ &  \\
\hline \hline
X-factor\footnote{Assumed X$_{\rm CO}$ from \citet{strong88}}\footnote{Mass inflow rate from \citet{Hatchfield2021}} & 2.23 $\pm$ 0.86 & 6.52 $\pm$ 2.53 & 2.3 $\pm$ 0.3 & 0.8 $\pm$ 0.6 & - \\
SED Fit & 0.13 $\pm$ 0.02 & 0.43 $\pm$ 0.07 & 0.15 $\pm$ 0.07 & 0.05 $\pm$ 0.03 & - \\
Max SED Fit & 0.32 $\pm$ 0.22 & 1.04 $\pm$ 0.72 & 0.37 $\pm$ 0.3 & 0.13 $\pm$ 0.11 & - \\
PPMAP & 0.28 $\pm$ 0.01 & 0.87 $\pm$ 0.04 & 0.31 $\pm$ 0.14 & 0.11 $\pm$ 0.05 & - \\
LTE $^{12}$CO\footnote{All LTE masses assume CO/H$_2 = 10^{-4}$} & 0.1 $\pm$ 0.04 & 0.3 $\pm$ 0.11 & 0.11 $\pm$ 0.06 & 0.04 $\pm$ 0.02 & - \\
LTE $^{13}$CO & 0.16 $\pm$ 0.08 & 0.48 $\pm$ 0.24 & 0.17 $\pm$ 0.11 & 0.06 $\pm$ 0.04 & $^{12}$C/$^{13}$C=25 \\
LTE C$^{18}$O & 0.14 $\pm$ 0.08 & 0.43 $\pm$ 0.24 & 0.15 $\pm$ 0.11 & 0.05 $\pm$ 0.04 & $^{16}$O/$^{18}$O=250 \\
LTE $^{12}$CO: $\tau$ Correction & 0.17 $\pm$ 0.06 & 0.51 $\pm$ 0.18 & 0.18 $\pm$ 0.1 & 0.06 $\pm$ 0.04 & $^{12}$C/$^{13}$C=25 \\
LTE $^{12}$CO: $\tau$ Correction & 0.25 $\pm$ 0.09 & 0.77 $\pm$ 0.28 & 0.27 $\pm$ 0.16 & 0.1 $\pm$ 0.06 & $^{12}$C/$^{13}$C=40 \\
LTE $^{12}$CO: $\tau$ Correction & 0.33 $\pm$ 0.12 & 1.01 $\pm$ 0.37 & 0.36 $\pm$ 0.21 & 0.12 $\pm$ 0.07 & $^{12}$C/$^{13}$C=53 \\
LTE $^{12}$CO: $\tau$ Correction & 0.48 $\pm$ 0.17 & 1.45 $\pm$ 0.53 & 0.51 $\pm$ 0.3 & 0.18 $\pm$ 0.11 & $^{12}$C/$^{13}$C=77 \\
LTE $^{12}$CO: $\tau$ Correction & 0.55 $\pm$ 0.2 & 1.67 $\pm$ 0.61 & 0.59 $\pm$ 0.34 & 0.21 $\pm$ 0.12 & $^{12}$C/$^{13}$C=89 \\
\hline
\end{tabular}
\label{tbl:massestimate}
\end{table*}

If we assume that G5 is a cloud flowing in towards the CMZ along the nearside bar lane, then it represents mass flowing along the bar. By measuring the mass in this cloud and comparing it to the \ce{CO} emission, we can reassess the measurement of how much mass is accreting onto the GC, with the caveat that G5 may not be representative of other clouds along the bar lane as the site of a collision. We chose to measure the mass using solely Field 2 of G5, as Field 1 has poor baseline fitting of \ce{^12CO} and is contaminated by emission from an HII region along the same line of sight. Throughout this section, we assume that the \ce{^12CO} to \ce{H2} abundance ratio is \SI{1e-4}{}. 



\subsubsection{CO-to-\ce{H2} X-Factor}

We first estimate the amount of mass in G5 by using the \citet{strong88} \ce{CO}-to-\ce{H2} conversion factor of \SI{2.3e20}{cm^{-2}~ (K~\kms)^{-1}}. This is the most commonly used Galactic \ce{CO}-to-\ce{H2} X-factor (X$_{\rm CO}$). We use an integrated intensity map of \ce{^12CO} $J=2\shortrightarrow1$, multiply it by 0.8 to account for the difference in intensity between \ce{^12CO} $J=1\shortrightarrow0$ and \ce{^12CO} $J=2\shortrightarrow1$ \citep{Leroy2009}, and then multiply it by the X$_{\rm CO}$ to create a column density map of \ce{H2} for Field 2. We use the molecular weight per hydrogen molecule $\mu_{H_2}=2.8$ \citep{Kauffmann2008} to calculate the mass in each pixel, and then summed over all of the pixels in Field 2 to give a mass estimate for the field. We report the estimated mass in Table \ref{tbl:massestimate} for this and all subsequent methods.

\subsubsection{Dust SED}
\label{sec:dustsed}

\begin{figure}[h!]
    \centering
    \includegraphics[width=0.45\textwidth]{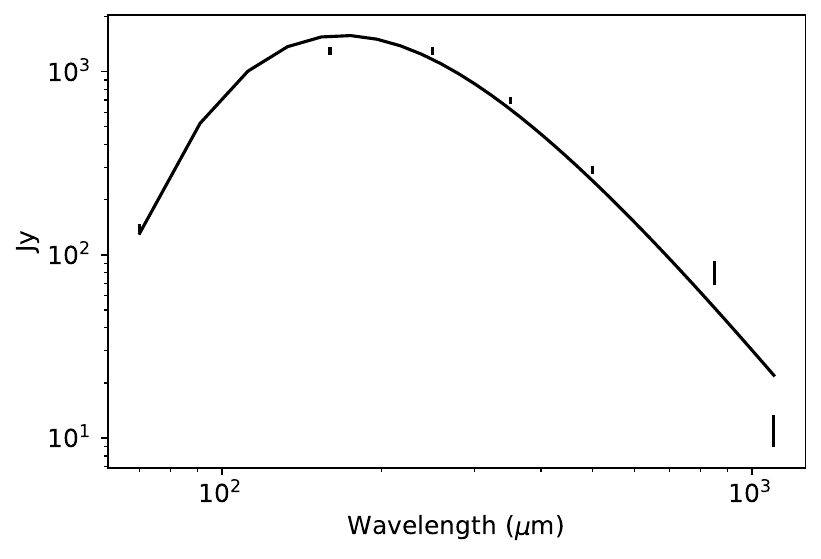}
    \caption{Dust emissivity SED fit of G5's Field 2 with a modified blackbody using \texttt{dust\_emissivity}. 
    We used Herschel SPIRE and PACS data from the Hi-GAL survey \citep{higal2016} for wavelengths 
    $\SI{70}{\micron}$, 
    $\SI{160}{\micron}$, 
    $\SI{250}{\micron}$, 
    $\SI{350}{\micron}$ and 
    $\SI{500}{\micron}$. 
    We used ATLASGAL for $\SI{850}{\micron}$ \citep{atlasgal2009} and BGPS for $\SI{1.1}{mm}$ \citep{bolocam2013}.
    The error bars are the quadrature sum of statistical uncertainty (background noise) and inherent uncertainty due to flux calibration. 
    The fit for the modified blackbody resulted in a 
    dust temperature of \SI{18.09}{K} $\pm$ \SI{1.19}{K}, 
    $\beta$ of \SI{1.75}{} $\pm$ \SI{0.29}{}, and 
    column density of \SI{4.34e21}{cm^{-2}} $\pm$ \SI{0.68e21}{cm^{-2}}.
    }
    \label{fig:dustsed}
\end{figure}

\begin{table}
\centering
\caption{SED Data}
\begin{tabular}{cccc}
\hline
Survey & Wavelength & Flux & Error \\
 & $\mathrm{\mu m}$ & $\mathrm{Jy}$ & $\mathrm{Jy}$ \\
 \hline
 \hline
PACS\footnote{PACS and Herschel \citet{higal2016}} & 70 & 139.7 & 7.0 \\
PACS & 160 & 1293.3 & 64.7 \\
Herschel & 250 & 1296.8 & 64.8 \\
Herschel & 350 & 696.3 & 34.8 \\
Herschel & 500 & 290.9 & 14.5 \\
ATLASGAL\footnote{\citet{atlasgal2009}} & 850 & 80.4 & 12.1 \\
BGPS\footnote{\citet{bolocam2013}} & 1100 & 11.1 & 2.2 \\
\hline
\end{tabular}
\label{tbl:sed}
\end{table}

We next estimate the amount of mass by creating a dust emissivity spectral energy distribution (SED) of Field 2, and then we fit the SED with a modified blackbody. 
Appendix Figure \ref{fig:dustshape} shows that the dust emission approximately matches the \ce{NH3} gas emission contours.
We used Herschel SPIRE and PACS data \citep{higal2016} for wavelengths 
$\SI{70}{\micron}$, 
$\SI{160}{\micron}$, 
$\SI{250}{\micron}$, 
$\SI{350}{\micron}$,  
$\SI{500}{\micron}$, 
ATLASGAL for $\SI{850}{\micron}$ \citep{atlasgal2009}, and BGPS for $\SI{1.1}{mm}$ \citep{bolocam2013}. 
We placed a rectangular aperture over Field 2 with a size of \SI{12.6}{\arcmin} by \SI{2.7}{\arcmin} for an area of \SI{33.7}{\arcmin} squared, 
placed a rectangular annulus around Field 2 with outer heights and widths twice that of Field 2, and subtracted the 10th percentile of the annulus from the aperture to remove background emission. 
We then summed over the annulus to find the background subtracted flux of Field 2.
We found the errors of each measurement by taking the quadrature sum of the statistical uncertainty of the measurements and the inherent uncertainty due to flux calibration. We then fit the values with a modified blackbody function. 
The measured values are shown in Table \ref{tbl:sed}. 
Figure \ref{fig:dustsed} shows the dust SED of Field 2.
The dust opacity is assumed to be defined by a continuous function of frequency defined by $\kappa = \kappa_0 (\frac{\nu}{\nu_0})^{\beta}$.
The modified blackbody used a value of $\kappa_0 = $ \SI{4.0}{cm^2 g^{-1}} at \SI{505}{GHz} \citep{battersby2011} and a gas to dust ratio of 100. 
The fit for the modified blackbody resulted in a dust temperature of \SI{18.1}{K} $\pm$ \SI{1.2}{K}, $\beta$ of \SI{1.8}{} $\pm$ \SI{0.3}{}, and \ce{H2} column density of \SI{4.34e21}{cm^{-2}} $\pm$ \SI{0.68e21}{cm^{-2}}. 
We expect that G5 has a dust temperature decoupled from the gas temperature, as the majority of the cloud has a lower volume density than the \SI{e6}{cm^{-3}} needed for collisional equilibrium \citep{clark2013}.

We also report an alternate fit of the Dust SED excluding the lowest wavelengths from the PACS survey. The modified blackbody fit of the remaining data resulted in a dust temperature of \SI{11.9}{K} $\pm$ \SI{3.4}{K}, $\beta$ of \SI{2.8}{} $\pm$ \SI{0.8}{}, and \ce{H2} column density of \SI{1.04e22}{cm^{-2}} $\pm$ \SI{7.2e21}{cm^{-2}}. This alternate fit of the data resulted in a column density measurement closer to that using the \citet{strong88} X$_{\rm CO}$ measurement, but uses less of the available data.

We then estimated the mass of G5's Field 2 using PPMAP \citep{ppmap2017}. 
PPMAP uses Hi-GAL data, Herschel PACS and SPIRE, to measure the dust column density by fitting a dust SED, using a factor of 100 dust to gas fraction.
PPMAP assumes a dust opacity value of $\kappa_0 = $ \SI{0.1}{cm^2 g^{-1}} at $\SI{300}{\micron}$, and $\beta = 2.0$. 
We made a cutout of Field 2 from the PPMAP column density map, multiplied each pixel by the pixel area, and then summed over the cutout to find the mass.

We estimated the total mass of G5 using PPMAP. We made a mask using the \ce{NH3} (3,3) data from the Mopra HOPS Survey \citep{hops3, hops1, hops2} to select only areas of the map with \ce{NH3} emission.
We identify those boundaries as the extent of G5, as shown in Figure \ref{fig:g5map}. We then applied the mask to PPMAP data, resulting in a total mass measurement of \SI{3.87e5}{M_{\odot}} $\pm$ \SI{0.45e5}{M_{\odot}}. This mass measurement does not account for dust along the same line-of-sight, such as from the overlapping HII region, and so is an upper estimate of the mass using this technique. 

\clearpage
\subsubsection{Local Thermodynamic Equilibrium}

\begin{figure}
    \centering
    \includegraphics[width=0.45\textwidth]{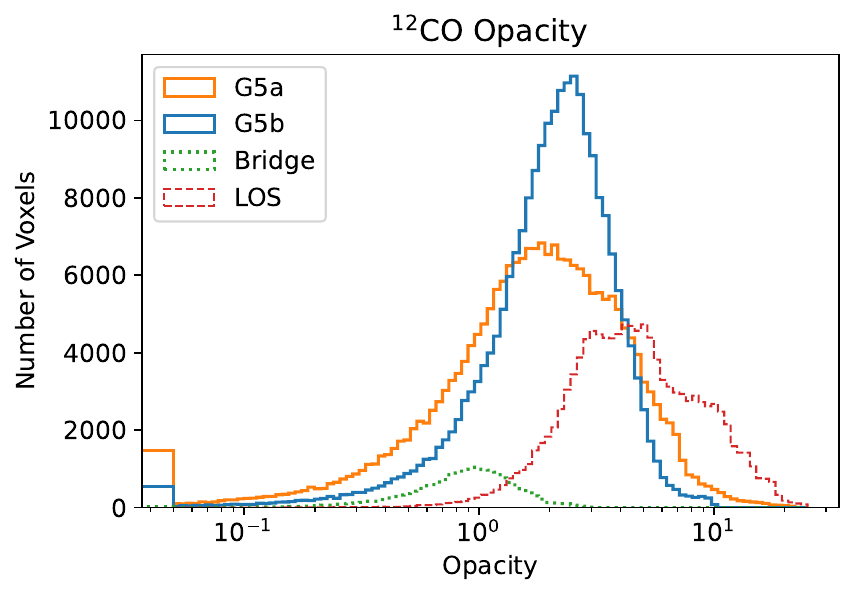}
    \caption{Histogram of the calculated \ce{^12CO} opacity of Field 2 assuming \ce{^12C}/\ce{^13C} = 40. Green shows the opacity values for voxels in the velocity bridge. Orange represents opacity values for the higher velocity cloud, G5a. Blue shows the opacity values for the lower velocity cloud G5b. Red represents the opacity values for the line of sight clouds between velocities of \SI{-50}{} to $\sim$ \SI{15}{\kms}.}
    \label{fig:opticaldepth}
\end{figure}

We estimated the mass of G5 using CO and its isotopologues by assuming they are in Local Thermodynamic Equilibrium (LTE). We note that if the \ce{CO} lines are subthermally excited, then the real column density of \ce{CO} found through this method is likely higher, but we find that this is unlikely. We used Equation 79 from \citet{mangum15} 
\begin{equation}
    \begin{split}
    N^{thin}_{tot} = \left(\frac{3h}{8 \pi^3 S \mu^2 R_i}\right)
    \left(\frac{Q_{rot}}{g_u}\right) 
    \frac{\exp\left(\frac{E_u}{k_B T_{ex}}\right)}
    {\exp\left(\frac{h \nu}{k_B T_{ex}}\right)-1} \\
    \times
    \int \frac{T_R d\nu}{f \left(J_\nu(T_{ex}) - J_\nu(T_{bg})\right)}
    \end{split}
    \label{eq:mangum}
\end{equation}
to calculate the column density where $h$ is Planck's constant, 
$Q_{rot}$ is the rotational partition function, $g_u$ is the degeneracy, 
$E_u$ is the energy of the upper energy level, 
$k_B$ is the Boltzmann constant, 
$\nu$ is the frequency of the transition, 
the sum of relative intensities 
$R_i = 1$ for $\Delta J = 1$ transitions, 
$f$ is the filling factor assumed to be 1, 
$T_{ex}$ is the excitation temperature, 
$\int T_R d\nu$ is the integrated intensity, 
$T_{bg}$ is the cosmic microwave background,
$J_\nu(T)$ is the planck function,
the line strength $S=\frac{J_u}{2J_u+1}$ for linear molecules where $J_u$ is the upper energy level, 
and the value for the molecular electric dipole moment ($\mu$) is from the Jet Propulsion Laboratory (JPL) Molecular Spectroscopy database and spectral line catalog \citep{Pickett1998}. 
We calculated the column densities for \ce{^12CO}, \ce{^13CO}, and \ce{C^18O}.  This equation assumes that the molecule being measured is optically thin.

While we can assume that \ce{^13CO} and \ce{C^18O} are optically thin in G5, we cannot assume the same for \ce{^12CO}. To remedy this, we estimate the optical depth of \ce{^12CO}. We can estimate the optical depth of \ce{CO} using Equation \ref{eq:taus_17},
\begin{equation} \label{eq:taus_17}
    \frac{I_{\nu} (^{12}CO)}{I_{\nu} (^{13}CO)} = \frac{1-e^{-\tau_{12}}}{1-e^{-\tau_{12}/R_C}}, R_C = \frac{\ce{^12C}}{\ce{^13C}}
\end{equation}
where $\mathrm{I_{\nu}}$ is the intensity of the lines and $\tau_{12}$ is opacity of \ce{^12CO}. The isotope abundance ratio \ce{^12C}/\ce{^13C} is lower in the Galactic Center than in the disk and is thought to increase radially outward from the Galactic Center \citep{Langer1990} due to the Galaxy forming from the inside out \citep{Chiappini2001, Pilkington2012}. 
\ce{^12C} is formed by He burning in massive stars on short timescales \citep{Timmes1995}, while \ce{^13C} is formed from \ce{^12C} seed nuclei in the CNO cycle of evolved low- and intermediate mass stars \citep{henkel1994}.

We estimate the opacity of \ce{^12CO} and the mass of Field 2 using Galactic abundances of  \ce{^12C}/\ce{^13C}. 
We use a root finding algorithm to solve to solve Equation \ref{eq:taus_17} for the opacity of \ce{^12CO} $\tau_{12}$. 
We spectrally and spatially reproject the \ce{^12CO} cube to the same spectral axis as the \ce{^13CO} cube and then divide the two for a ratio cube of \ce{^12C}/\ce{^13C}. We then use root finding to solve Equation \ref{eq:taus_17} to estimate the optical depth of every voxel in the cube. 
We find separate optical depth cubes for the Galactic values of \ce{^12C}/\ce{^13C} reported in \citet{Henkel1985, wilson94, Riquelme2010}. 
Bars radially mix gas, so the \ce{^12C}/\ce{^13C} ratio may not be much different from the Galactic Center.
\citet{Riquelme2010} measures conflicting \ce{^12C}/\ce{^13C} ratios at G5, resulting in one measurement higher than 70 for a \SI{87}{\kms} component using \ce{H^12CO+}/\ce{H^13CO+}, and other measurements as low as 10 to 45 using other components and line ratios.
We set the opacity values to 0 for each voxel with a value of \ce{^12CO} or \ce{^13CO} less than one standard deviation, and we set the opacity to 0 for \ce{^12CO}/\ce{^13CO} ratios less than 1 or greater than the assumed \ce{^12C}/\ce{^13C}.
We show the distribution of calculated opacity values for each voxel in the opacity cube assuming \ce{^12C}/\ce{^13C} = 40 in Figure \ref{fig:opticaldepth}.


We then apply each estimated optical depth cube to the \ce{^12CO} cube using the linear relationship between the integrated intensity and total column density using Equation 86 from \citet{mangum15}, 
\begin{equation}
    N_{tot} = N^{thin}_{tot}
    \frac{\tau}{1-exp(-\tau)}
    \label{eq:tau_factor}
\end{equation} 
where $N_{tot}$ is the total column density taking into account optical depth, $\tau$ is the optical depth of \ce{^12CO}, and $N_{tot}^{thin}$ is the estimated LTE column density assuming the line is optically thin from Equation \ref{eq:taus_17}. We then take the integrated intensity of the altered cubes and solve for the column density and mass of the field.

Finally, we estimate the mass of G5 using the optically thin \ce{CO} isotopologues \ce{^13CO} and \ce{C^18O}. The average column densities and the calculated X$_{\rm CO}$ for each above method are included in Table \ref{tbl:massestimate}. The calculated X$_{\rm CO}$ uses the average velocity integrated intensity of \ce{^12CO} over Field 2 adjusted to account for the difference intensity between \ce{^12CO} $J=1\shortrightarrow0$ and \ce{^12CO} $J=2\shortrightarrow1$ \citep{Leroy2009}, \SI{283.46}{K~\kms}.

\section{Discussion} 

\begin{figure}[t]
    \centering
    \includegraphics[width=.5\textwidth]{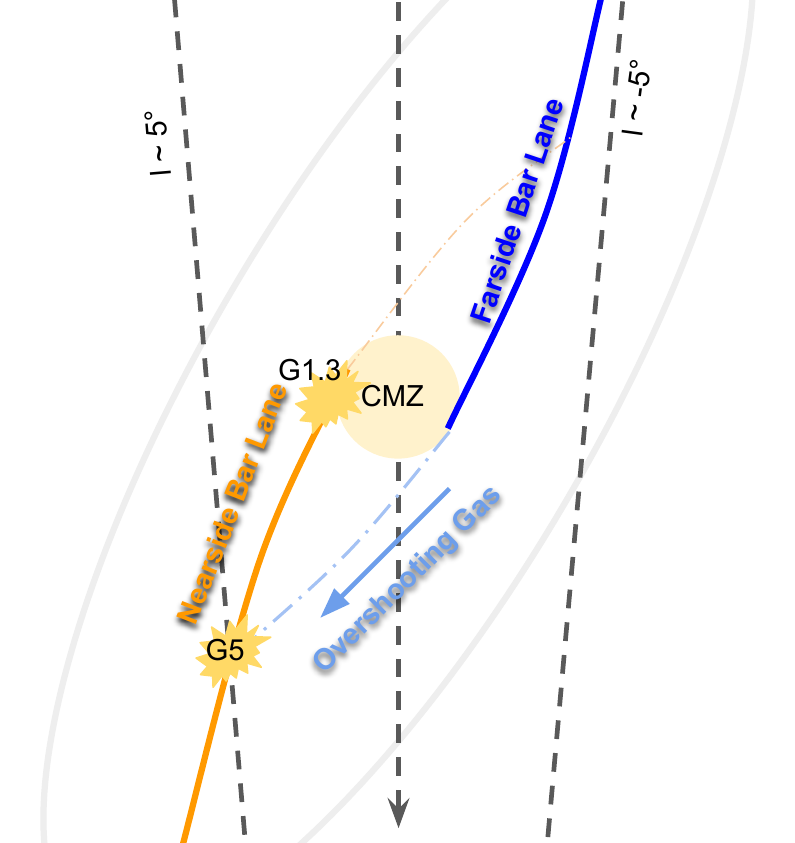}
    \caption{Cartoon of a top down view of the Milky Way's bar. Not to scale. 
    Two bar lanes bring material falling from X$_1$ orbits to X$_2$ orbits.
    The yellow burst markers are some of the sites where cloud-cloud collisions are thought to happen between streams of gas. 
    G5 is the likely site of a collision between overshooting gas from the far side bar lane and the nearside bar lane. G1.3 is the likely site of gas accreting onto the CMZ from the nearside bar lane \citep{Busch2022}.  
    The dashed lines represent lines-of-sight from positions in the center of the Galaxy and along the bar to the observer's position. The grey ellipse represents X$_1$ orbits. The yellow circle represents X$_2$ orbits, which could be the CMZ. 
    \label{fig:cartoon}}
\end{figure}

\begin{figure*}[ht]
    \centering
    \includegraphics[width=.45\textwidth]{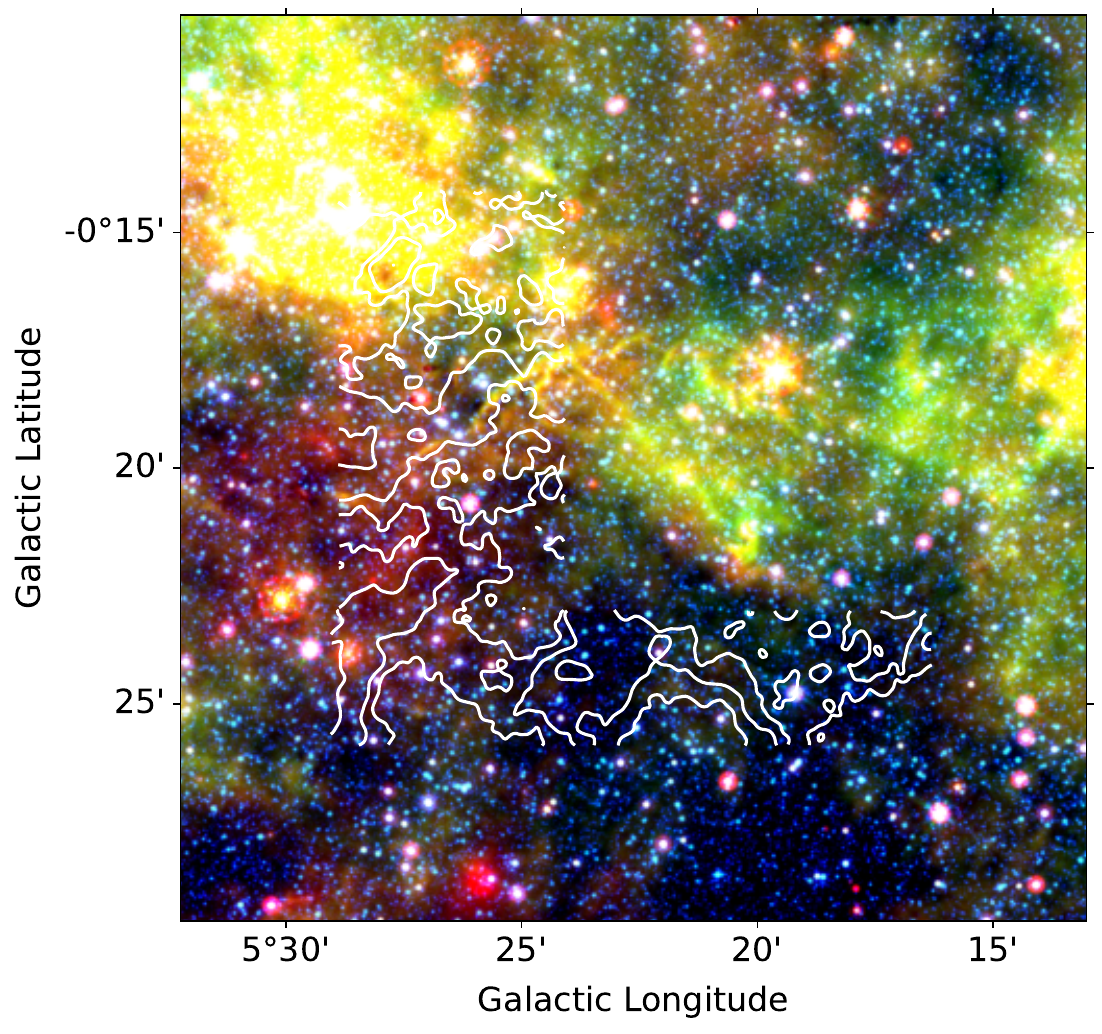}
    \includegraphics[width=.45\textwidth]{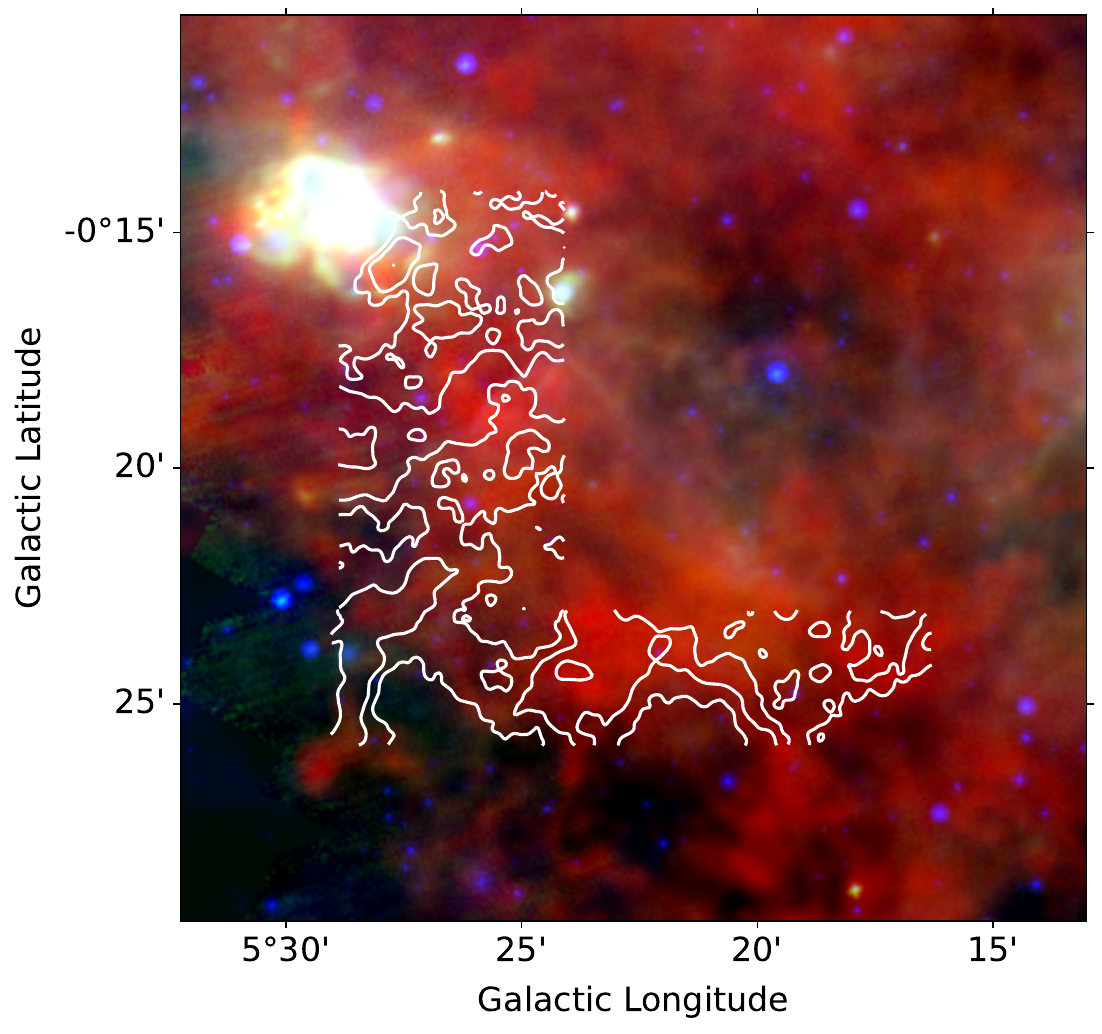}
    \caption{
    Wider context images of G5 showing the projected environment around the cloud using Herschel and Spitzer data. 
    The HII region in the field is not associated with G5. 
    The white contours are from the integrated intensity map of H$_2$CO (3$_{03}-2_{02}$), and do not seem to follow the morphology of the images. 
    Left: Three color Spitzer image of MIPSGAL 24 $\mu$m, 8.0 $\mu$m and 4.5 $\mu$m. 
    Right: Three color image of Herschel SPIRE 250 $\mu$m and PACS 70 $\mu$m, and MIPSGAL 24 $\mu$m.
    }
    \label{fig:context}
\end{figure*}

We have shown that G5 is the site of warm, shocked, dense gas with a PV diagram that includes a velocity bridge. We now go on to interpret the meaning of these features. In Section \ref{sec:velocitybridge}, we will talk about how the velocity features of G5 indicate a cloud-cloud collision. Section \ref{sec:discussgastemp} discusses the warm temperatures detected in G5 and possible causes of heating in the cloud. Next, in Section \ref{sec:collisionenergy} we will discuss the energy of the collision in G5 and how such a large collision lacks strong shock tracers. Then, Section \ref{sec:overestimatemass} will speak on the inconsistencies in mass estimate methods, and Section \ref{sec:accretionrate} will discuss the numerical implications for Galactic mass inflow estimates. Finally, in Section \ref{sec:asymmetric}, we will clarify the distinction between G5 and Bania 1 as a pair of clouds outside of the CMZ.

\subsection{Properties of the Cloud}
\subsubsection{Velocity Features}
\label{sec:velocitybridge}

We begin to see evidence of a cloud-cloud collision in Figure \ref{fig:separate}, where integrating the intensity over two different ranges of velocities shows distinctly different structures. The velocity map in Figure \ref{fig:mom12} better displays the vastly different velocities of G5a and G5b, and the gradient between them that starts in the middle of Field 2 as red transitions to white where there is a peak in emission and then swaps fully to blue within a few arcminutes. This alone does not mean that there is a cloud-cloud collision. Molecular clouds may overlap along a line of sight. 
The velocity dispersion map in Figure \ref{fig:mom12} finds an extremely wide line width in the same region of the gradient in the velocity map. 
However, the elevated dispersion is due to overlapping spectral features, as shown in the second panel of Figure \ref{fig:spectra}. Further evidence is needed to confirm the cloud-cloud collision. To that end, we view the region of the suspected collision in position-velocity space.

Figure \ref{fig:pvdiagram} shows a PV diagram of Field 2 where the offset is in Galactic Longitude. 
The high velocity component in the PV diagram is clearly the same high velocity feature in the velocity field map in Figure \ref{fig:mom12} identified as G5a, and the low velocity component is G5b, the low velocity spectral feature in blue in the same map. 
Visible in PV space 
is a wide spectral feature stretching between G5a and G5b. This spectral feature is narrow in position space, as it is only about an arcminute across at its widest, but it has a length of $\sim$\SI{100}{\kms}. This feature is a velocity bridge.

A velocity bridge is a feature in a PV diagram which is wide in velocity space but relatively narrow in position space, and connects two features. A velocity bridge indicates that the two clouds are interacting, instead of being coincidentally along the same line-of-sight where they overlap \citep{HaworthDec15, HaworthJun15}. As shown in Figure \ref{fig:pvdiagram}, a velocity bridge is clearly visible in the center PV diagram of Field 2. 
When two clouds collide, only a small amount of mass is involved in the collision at a time. Gas fills the entire space between the clouds, with different velocities in the bridge corresponding to different amounts of either cloud. 
Velocity bridges tend to last as long as the crossing time of the cloud, but for streams of gas flowing along bar lanes, the velocity bridge may remain for longer times. 
The simulation from \citet{sormani19} suggests that streams of gas overshooting the CMZ collide with material along bar lanes on the other side of the Galaxy certain at certain locations with vastly different line of sight velocities, a cartoon representation of which is shown in Figure \ref{fig:cartoon}. 
This simulation resembles the observed velocity bridge and the $\sim$ \SI{100}{\kms} velocity difference between G5a and G5b where they collide. 

A cloud complex that can be compared to G5 is G1.3, which has a velocity bridge identified in \ce{CS} $J=2\shortrightarrow1$ \citep{Busch2022}.

\subsubsection{Gas Temperature}
\label{sec:discussgastemp}

The gas temperatures of G5a and G5b shown in Figure \ref{fig:temphist} are comparable to typical Galactic Center temperatures at an average of \SI{60}{K}, but less than the more extreme temperatures \citep{ginsburg16, Krieger2017}. The temperatures of G5a and G5b are comparable to each other, but a KS test of the data shows that the distribution of temperatures from the clouds are unlikely to be the same.
 
The gas temperature measured using \ce{H2CO} line ratios is much higher than the \SI{18.1}{K} measured from the dust emissivity SED model of Field 2 in Figure \ref{fig:dustsed}. 
The gas and dust temperatures are de-coupled, meaning that the gas is not efficiently cooled by interactions with the dust, likely due to G5 having a volume density too low to couple with dust \citep{clark2013, ginsburg16}.

\citet{akhter21} found the gas rotational temperature of G5 using a transition line ratio of \ce{NH3} (2,2)/(1,1), finding gas temperatures between \SI{60}{K} to \SI{100}{K}. 
These temperatures are consistent with the range of gas temperatures found with \ce{H2CO} line ratios.

Heating in molecular clouds has several different causes: radiation, cosmic rays, and shocks.

First, we rule out radiation as a heating mechanism in G5.
We looked at Herschel and Spitzer data sets of G5, the three color images shown in Figure \ref{fig:context}. While Figure \ref{fig:context} shows the presence of an HII region at the top of Field 1, that HII region is not associated with G5. \citet{wink83} found the velocity of the HII region using \ce{H76 \alpha}, resulting in a measurement of \SI{28.3}{\kms} \SI{\pm 0.9}{\kms} with a FWHM of \SI{20.8}{\kms} \SI{\pm 2.0}{\kms}. While \ce{CO} emission from the HII region is observed in Field 1, the emission is at a different velocity compared to the rest of the material associated with G5 nearby. There are no features which resemble stars interacting with gas in G5. There is no evidence of feedback from star formation, so we do not expect radiation to be the primary source of heating in G5.

Cosmic rays are also unlikely sources of heating in G5.
While the Galactic center has a higher cosmic ray ionization rate (CRIR) than the Galactic disk \citep[CRIR$\sim10^{-14}~\mathrm{s}^{-1}$;][]{Yusef-Zadeh2007,Indriolo2015,oka2019}, there is no evidence for such elevated CRIR in the Galactic bar.
Unless there were some local source of cosmic rays, we would not expect CRs to heat the molecular gas to Galactic center temperatures.
We also do not see evidence of any such source, such as a supernova remnant, near G5.

We find that the most likely heating mechanism in G5 is shocks.
Since the clouds are colliding, we expect shocks to be a significant heating mechanism in the molecular gas.
\citet{ginsburg16} identifies turbulent dissipation (shocks) as one of the key driving processes behind the high gas temperatures measured in the Galactic Center. 
We measure very wide line widths in G5, as shown by Figure \ref{fig:mom12} and the Position-Velocity diagrams in Figures \ref{fig:pvdiagram} and \ref{fig:all_pvdiagrams}. 
The collision between G5a and G5b causes an increase in the kinetic temperature as gas from vastly different velocities collides and mixes together. 
While the collision may explain high temperatures found near the location of the cloud-cloud collision, the entirety of the clouds, not just the bridge feature, are warm. 
\citet{sormani21} finds that a cloud-cloud collision such as the one found in G5 would not happen as one large collision of gas decelerating from \SI{200}{} to \SI{0}{\kms}, but as a series of much weaker shocks that allow the gas to cool as it is being shocked over time. 
Another form of shock heating is tidal shear stress heating, which would be experienced by material in bar lanes as it is stretched and manipulated by the gravity of the Galaxy's bar potential. 
Tidal stress may also cause the \ce{SiO} $J = 5\shortrightarrow4$ observed in G5.

We expect that shocks from tidal shear and the cloud collision cause turbulence within G5, warming it up to the temperatures we observe. G5 has likely been the location of many previous collisions, as evidenced by the `spur' feature in the PV diagrams (Appendix Fig. \ref{fig:labelled_pv}), which would heat the cloud repeatedly over time.
\citet{ginsburg16} shows in Appendix F that line widths of $\sim$ \SI{20}{\kms} and temperatures from \SI{40}{} to \SI{60}{K} match well with \texttt{DESPOTIC} models for turbulent heating at a volume density of \SI{e5}{cm^{-3}}. 
Shocks are able to explain the observed gas temperature.

\subsubsection{Energy of the Collision}
\label{sec:collisionenergy}

The cloud-cloud collision in G5 happens between gas with a velocity difference of over \SI{100}{\kms} along the line of sight. As the clouds collide on an angle relative to the line of sight, we only see a component of the velocity of the collision, so the true velocity of the collision is even higher. A large amount of energy must be involved in this collision.

Using a lower mass estimate of \SI{e4}{M_{\odot}} from Section \ref{sec:massest} for the mass involved in the collision, with a velocity difference of over \SI{100}{\kms}, the collision would produce $\sim$ \SI{e51}{erg} of kinetic energy, roughly equivalent to the mechanical energy from a supernova. We expect heating of up to \SI{e5}{K} from the J shock due to the collision between the clouds, which would produce weak X-rays and ionize atoms in the region of the shock. Such a large collision would disassociate molecules, but we detect \ce{CO} in the velocity bridge, the gas directly involved in the collision. There are a few options for where the energy of the collision is going. The first is that there is one big shock that is heating the gas up to \SI{e5}{K}, but the molecules reform soon after the shock. The second is that the collision has only just begun and involves only the less dense, outer layers of the cloud. The third possibility is that molecules are not being destroyed. Instead of one big shock, the collision happens as a series of smaller shocks, which would heat up the gas much less \citep{sormani21}.

Some of the energy goes into increasing the thermal energy of the clouds involved in the collision. 
In Section \ref{sec:discussgastemp}, we measure the kinetic temperature of the cloud's gas at approximately \SI{60}{K}. 
The internal kinetic energy of the clouds can be estimated using the velocity dispersion, and is $\sim$ \SI{e50}{erg}. 
The internal energy of G5 is ten times less than the collisional energy.

Much of the collision energy is radiated away by dust.
The length of the collision is approximately the crossing time. Assuming the crossing length is \SI{20}{pc} and the clouds collide with a constant velocity of \SI{100}{\kms}, the crossing time is \SI{200}{kyr}. The energy rate for $\mathrm{M_{cloud}v_((collision))^2 \sim}$ \SI{e51}{erg} over \SI{200}{kyr} is \SI{e5}{L_{\odot}}. 
The dust luminosity can be estimated using the dust SED fit from Section \ref{sec:dustsed}. The luminosity of dust with a temperature of \SI{18}{K} in an area of \SI{16}{pc^2} is \SI{2e5}{L_{\odot}}. 
We assume that the dust in G5 has a temperature in equilibrium, although the temperature we measure is higher than the solar neighborhood. 
We do not know what the interstellar radiation field or cosmic ray rate is at G5, but if they are stronger then we expect them to affect the equilibrium temperature of the dust to some extent. 
The energy generated by the collision is approximately the same as the amount of energy radiated by the dust.

The G5 cloud-cloud collision should
produce a strong signal in shock tracers such as \ce{SiO}. 
\ce{SiO} is a strong shock tracer, often seen in strongly shocked regions of the ISM such as in molecular outflows \citep{schilke97}.
We lack a detection of \ce{SiO} $J=5\shortrightarrow4$ in the gas of the velocity bridge, which is directly involved in the collision. In Section \ref{sec:shocks}, we simulate the line in non-LTE using \texttt{DESPOTIC}, finding that a low volume density in the velocity bridge material would prevent the detection of \ce{SiO} $J=5\shortrightarrow4$ in the velocity bridge. The excitation we see in the main bodies of G5a and G5b are consistent with the presence of shocks in the clouds, but there is no evidence of a high velocity, strong shock. 
A possible cause for the lack of evidence of a strong shock from the collision is that G5a and G5b are colliding not as monolithic masses, but as a series of weaker shocks, allowing the heat to dissipate more efficiently than two large masses colliding at once \citep{sormani21}. These weak shocks would not destroy the dust grains as efficiently as a strong shock, resulting in a smaller \ce{SiO} abundance than expected from the collision.

\subsubsection{X-Factor Overestimates Bar Lane Gas}
\label{sec:overestimatemass}

Initial estimates of the optical depth of G5 suggest that the gas flowing into the CMZ is not very optically thick. 
The \citet{strong88} X$_{\rm CO}$ relies on the \ce{CO} being optically thick, but smaller optical depths measured in Figure \ref{fig:opticaldepth} would result in an overestimation of the mass when using the X$_{\rm CO}$. 
Comparing mass estimates using the X$_{\rm CO}$ and dust mass estimates from Table \ref{tbl:massestimate} suggest that the mass of G5 is $\sim$10-20x overestimated by the accepted X$_{\rm CO}$. 
\citet{Liszt2006} measured a mass of \SI{6.1e6}{M_{\odot}} for G5 over an area of \SI{180}{pc} x \SI{60}{pc} using observed \ce{CO} $J=1\shortrightarrow0$ and an X$_{\rm CO} = \SI{2e20}{cm^{-2}~ (K~\kms)^{-1}}$. 
We measured \SI{3.87e5}{M_{\odot}} $\pm$ \SI{0.45e5}{M_{\odot}} for the total mass of G5 using PPMAP column density measurements \citep{ppmap2017}, for a 16x difference between the two measurements. 
The consequences of this difference affect the results of \citet{sormanibarnes19} and other Galactic mass flow estimates, which overestimate of the amount of the CMZ mass inflow rate. 

We measure \SI{1.5e19}{N(H_2) ~ cm^{-2} (K~\kms)^{-1}} as the X$_{\rm CO}$ in G5. \citet{ferriere2007} find that X$_{\rm CO}$ near the Galactic Center is of the order of $\sim$\SI{e19}{N(H_2) ~ cm^{-2} (K~\kms)^{-1}}. 
We calculated that G5 is \SI{1.33}{kpc} distance from the Galactic Center, meaning that the lower value of X$_{\rm CO}$ extends along the Galactic bar.
The lower opacity of \ce{^12CO} in G5 is likely due to the wide velocity dispersion of the cloud. If bar dynamics make the opacity and hence the X$_{\rm CO}$ different, 
then there is a significant impact on bar mass inflow rates calculated using the standard Galactic X$_{\rm CO}$. 
A caveat of our estimate is that G5 might be special due to being the site of a collision, so it may not fully represent all of the gas on the bar lanes.

\subsection{Geometry of the Bar}

The Milky Way is a barred spiral galaxy. In the bar, material is expected to flow along two main types of orbits, X$_1$ and X$_2$ \citep{contopoulos1989}. X$_1$ orbits are nested, elongated ellipses that make up the main body of the bar, but become self intersecting as material loses angular momentum due to crossing orbits \citep{anthanassoula1992}. 
\citet{sormani19}'s model is a hydrodynamic simulation of the Galactic bar showing material flowing along bar lanes towards the inner ring of the CMZ. Some of that material is shown overshooting the CMZ, continuing along a trajectory to collide with the bar lane on the opposite side of the Galaxy. A cartoon depiction of the path of the overshooting gas is shown in Figure \ref{fig:cartoon}. 

\citet{sormani19} and \citet{Liszt2006} identify several ``Extended Velocity Features" (EVF) using \ce{^12CO} $J=1\shortrightarrow0$ from \citet{bitran97}, one of which is G5. 
In Figure 1 of \citet{sormani19}, G5 is identified in green as an EVF at $\ell=5.4^{\circ}$. 
Figure 3 of the paper and Figure \ref{fig:pvdiagram}, feature V1 looks similar in ($\ell$, $v$) space to the EVF of G5. 
Figure 4 of the same paper shows the simulated line of sight velocities of the bar from a top-down view of the bar. In the larger of the circles, the one marking where overshooting gas hits the bar lane, the line of sight velocities are a mix of lower velocity and higher velocity gas, exactly the same as G5. 
G5 is strong observational evidence for gas overshooting the CMZ and crashing into the bar lane on the opposite side of the Galaxy.

As G5a is gas that is traveling down a bar lane, the properties of the cloud tell us about the initial properties of gas as it enters the CMZ. 
The CMZ contains roughly 5\% of the molecular gas mass of the Milky Way \citep{Henshaw2022}, but lacks the amount of star formation that would be expected based on the large amount of very dense gas. 
The warm temperatures of the gas in G5 suggests that the warm temperatures observed in gas in the CMZ is not solely due to the extreme radiation and cosmic rays that come from the abundance of star formation. 
G5 contains no massive star formation, so any heating must be due to shocks.

A velocity bridge at G1.3 was detected in \ce{CS} $J=2\shortrightarrow1$, likely representing gas further along the nearside bar lane accreting onto the CMZ and interacting with gas already present \citep{Busch2022}. 
G1.3 and G5 are both cloud complexes along the nearside Galactic bar lane that show cloud interactions along a line of sight with high velocity differences.

\subsubsection{Accretion Rate onto CMZ}
\label{sec:accretionrate}

\citet{sormanibarnes19} find that the mass inflow rate along the near side bar lane is \SI{1.2}{\textup{M}_\odot~ yr^{-1}}, along the far side is \SI{1.5}{\textup{M}_\odot~ yr^{-1}}, for a combined total of \SI{2.7}{\textup{M}_\odot ~ yr^{-1}} flowing along bar lanes towards the CMZ. 
Updated estimates find that not all of the gas in the bar lane accretes onto the CMZ immediately, as only about 30\% of the gas flowing along bar lanes accretes onto the CMZ, accreting with a rate of only \SI{0.8}{M_{\odot}~ yr^{-1}} \citep{Hatchfield2021}.
These mass inflow rates were found using the \citet{strong88} X$_{\mathrm{CO}}$ of \SI{2.3e20}{cm^{-2} (K~\kms)^{-1}} to estimate the amount of mass flowing into the CMZ, but we find that accepted X$_{\mathrm{CO}}$ overestimates the amount of mass in G5 in Section \ref{sec:massest}.

Mass inflow rates using the measured X$_{\rm CO}$ in G5 are lower than previously measured rates. 
Table \ref{tbl:massestimate} includes estimates of the CMZ mass inflow rate adapted from measurements by \citet{Hatchfield2021}, using the X$_{\rm CO}$ calculated using estimates of the column density and \SI{283.5}{K~\kms} for the average measured integrated intensity of \ce{^12CO} $1\shortrightarrow0$. 
Our calculated mass inflow rates are consistently less than the current value of \SI{0.8}{\textup{M}_\odot ~ yr^{-1}} \citep{Hatchfield2021}. 
The calculated mass inflow rates are close to the observed star formation rate of the CMZ. 
The rates are also less than the mass flowing out of the CMZ through Fermi bubbles and other outflows of \SI{0.5}{\textup{M}_\odot ~ yr^{-1}} \citep{Bordoloi2017, Teodoro2018, Teodoro2020}.

The most likely cause of the lower X$_{\rm CO}$ observed in G5 is due to the low optical depth of the region. 
Figure \ref{fig:opticaldepth} shows the opacity values calculated for Field 2 assuming \ce{^12C}/\ce{^13C} = 40. 
The velocity bridge, show in green, contains gas immediately involved in the observed cloud-cloud collision and seems optically thin as its opacity values drop off after 1. 
The two colliding clouds G5a and G5b are only moderately optically thick, with their opacity distributions peaking at around 2 and 3 respectively. 
The line of sight clouds, assumed to be somewhere else in the Milky Way but along the same line of sight as G5, show much higher opacity values, very few being under $\tau=1$.
The \citet{strong88} X$_{\rm CO}$ assumes that \ce{CO} is optically thick, so if all of the gas along the bar has an optical depth like G5, then the accepted measured mass flow into the CMZ is an overestimation.

Lower values of X$_{\rm CO}$ have been observed in the bars and bar lanes of other spiral galaxies \citep[e.g.][Teng et al submitted, Sormani et al submitted]{meier2001, Bolatto2013, teng2022}. Our results are consistent with, and offer additional explanation to these results.

%

\subsubsection{G5 and Bania 1 are Not Symmetric Partners}
\label{sec:asymmetric}

\citet{akhter21} suggest that G5 and B1 are an anti-symmetric pair, meaning that they have the same position in Galactic coordinates but with the signs flipped. They claim that the two clouds, due to their similar angular distances from the Galactic Center and ammonia parameters, are related and trace identical features of the Galactic bar. They report that the clouds host hot gas and have wide emission lines due to shock heating. 
They offer two possibilities for the identity of the clouds: 
(1) the clouds are at the leading edges of the Galactic bar, having possibly passed through the bar lane shocks or
(2) the clouds are on the innermost X$_1$ orbit, and collide with the gas where the orbits become self-intersecting. 

However, in our model, the anti-symmetric position of G5 and B1 with respect to the Galactic center is likely coincidence. Figure \ref{fig:geometry} shows a model of the geometry of the Galactic bar, where the solid green lines point to the two ends of the bar and the dashed dark blue line points to the Galactic center. As shown in Figure \ref{fig:geometry}, the angle between the line of sight to the near end of the bar and to the Galactic center is larger than the angle between the far end of the bar and the line of sight to the Galactic center. The dashed green line has the same angle to the line of sight to the Galactic Center as the line of sight to the near end of the bar. As shown in the figure, the dashed green line does not point towards a symmetric feature on the far end of the bar. The foreshortening of the closer end of the Galactic bar means that it will appear at a larger angle from the Galactic center than the farther end of the bar. Therefore, while they may both be along the Galactic bar, G5 and B1 cannot both trace the ends of the bar nor be symmetric features of the Galaxy unless the bar was not tilted. 
As shown in Figure \ref{fig:cartoon}, G5 is likely somewhere along the length of the nearside bar lane. Assuming that the cloud is on the bar, B1's geometry places it further out from the Galactic Center than G5, perhaps at the end of the \SI{135}{\kms} arm \citep{fux1999}. 

\subsubsection{Magnetic Loops}
An alternative explanation for the EVFs \citep{Liszt2006, sormani19} along the Galactic plane are the foot points of magnetic loops. \citet{fukui2006} identified two loop-like structures extending vertically out of the Galactic plane, which were concluded to be material lifted out of the Galactic plane by magnetic buoyancy, analogous to solar loops caused by the Parker instability.

According to the model, in a strong enough magnetic field, a small perturbation in the field causes the magnetic field lines to pinch and suddenly lift material out of the plane of the Galaxy. Gravity pulls the material back down along the slope of the magnetic perturbation towards the Galactic plane, creating a loop. Where that material intersects the Galactic plane is thought to be the foot point of the magnetic loop. \citet{fukui2006} associates these foot points with areas of warm, dense gas with wide velocity dispersion, exactly like EVFs. One of the foot points of the observed loops is B1 (Fig. \ref{fig:galoverview}), which has been compared to G5 due to their axisymmetric Galactic coordinates \citep{akhter21} and \ce{NH3} (3,3) emission features. G5 is one of the other hypothetical magnetic foot points, but on the other side of the Galaxy from B1.

We find that it is unlikely that G5 is evidence of the foot point of a magnetic loop. 
The positions of the clouds involved with the cloud-cloud collision at G5 are displaced in Galactic Longitude rather than in Galactic Latitude, as predicted by the magnetic loop foot point model.
The loop foot point model suggests that material falls down onto the plane of the Galaxy, so that that the collision would have a major component in the z-direction, or Galactic Latitude. 
As shown by observations from \citet{Enokiya2021} and \citet{Torii2010}, the foot points of magnetic loops have cloud-cloud collisions associated with velocity bridges identified with a cut along Galactic Latitude.
The cloud-cloud collision at G5 is distinctly different from these observed loops, as it is identified with a cut along Galactic Longitude, as shown by Figure \ref{fig:pvdiagram}, showing that the collision is likely happening within the plane of the Galaxy. 
G5 is not likely to be the foot point of a magnetic loop, as the collision does not involve gas that has risen out of the plane of the Galaxy.

Additionally, \citet{sormani19} shows that G5 is connected to bar lanes. The velocity of the G5 EVF stops exactly at the velocity of the bar lane, and G5 is spatially associated with bar lane gas. Bar lanes have no role in the Parker instability model, so the association between G5 and the bar lanes is better explained by the explanation that G5 is due to overshooting gas.

We also find it unlikely that the magnetic field strength at G5 is strong enough for the Parker instability to lift material out of the Galactic plane. The Galactic Center has an estimated magnetic field strength of $\mathrm{\SI{100}{\mu m}}$ to \SI{1}{mG} \citep{Henshaw2022}. The reported magnetic field strength to induce the Parker instability is $\SI{150}{\mu G}$, but G5 is not in the Galactic Center and is not likely to have the necessary magnetic field strength. 
According to the \citet{suzuki2015} model, the magnetic field at the Galactic radius of G5 is only $\sim$ $\mathrm{\SI{10}{\mu G}}$, which is too low for the Parker instability.

\section{Conclusion}

We observed G5 at $(\ell,b) = (+5.4, -0.4)$ using ALMA/ACA. We observed \ce{^12CO} $2\shortrightarrow1$, \ce{^13CO} $2\shortrightarrow1$, \ce{C^18O} $2\shortrightarrow1$, \ce{H2CO} $3_{2,1}\shortrightarrow2_{2,0}$, \ce{H2CO} $3_{0,3}\shortrightarrow2_{0,3}$, and \ce{SiO} $5\shortrightarrow4$.

G5 is strong observational evidence of gas overshooting the CMZ and entering different orbits to collide with a bar lane on the other side of the Galaxy. We observed two velocity features connected with a velocity bridge in a PV Diagram of Field 2. We conclude that G5 is comprised of two colliding clouds, which we called G5a and G5b. 
Based on \citet{sormani19}'s model, G5 is the location of a high velocity cloud-cloud collision along a bar lane feeding material into the CMZ. 
The model suggests that the cloud-cloud collision at G5 is made of a cloud (G5b) which has flowed down the bar lane on the other side of the Galaxy, overshot the CMZ, and collided with the bar lane on this side of the Galaxy (G5a).
This is one of the highest velocity cloud-cloud collisions in the Galaxy. 
The results have many implications for our understanding of how the inner Galaxy is structured, and how the flow of material into the CMZ works.

Gas flowing into the CMZ is warm as it falls into the Galactic Center. We measured the temperature of G5 using \ce{H2CO} ratios of order \SI{60}{K}. This measured temperature shows that G5 is warmer than Galactic disk, but comparable to Galactic Center molecular clouds.
We observed a lack of ongoing massive star formation associated with the cloud cloud collision in existing data sets in the infrared. This lack of star formation in G5 implies that the gas is heated by shocks through the cloud collision between G5a and G5b, as well as tidal shear stress heating of the gas as G5 closely orbits the Galactic Center.

We observed that the \ce{^12CO}/\ce{^13CO} ratio in G5 is consistent with optically thin, or at most marginally optically thick \ce{^12CO}. 
We re-measured the local X$_{\rm CO}$ factor by comparing the \ce{^12CO} integrated intensity to the dust-inferred total mass. 
We found X$_{\rm CO}$ = \SI{0.15e20}{N(H_2) ~ cm^{-2} (K~\kms)^{-1}} fitting the dust emissivity SED and 
\SI{0.31e20}{N(H_2) ~ cm^{-2} (K~\kms)^{-1}} using PPMAP, which is 10-20x less than the typical \citet{strong88} X$_{\rm CO}$. 
Using this X$_{\rm CO}$, we re-measure the gas inflow rate, finding it is 10-20x lower than reported in \citet{sormanibarnes19}, although G5 might be extreme in that it is a cloud collision site and may not represent the average cloud along the bar lanes.

We observed G5, which shows strong evidence for the inflow of gas along the Galactic bar and overshooting the CMZ. We have revealed that G5 is the location of a cloud cloud collision, has a temperature of around \SI{50}{K}, and has a lower optical depth than expected of a GMC. There remain open questions about the origin of G5 and chemistry of the gas. 
G5 and similar clouds could be used as laboratories to study what happens when two molecular clouds smash together at high velocities.
Since the typical X$_{\rm CO}$ does not apply to G5, in the future we must treat gas along bars with additional care when estimating the masses.

\acknowledgments
SRG thanks the National Radio Astronomical Observatory the two summer internships which started this research. 
SRG would like to thank Dr. Desika Narayanan for useful discussions on the impact of large line widths on optical depth measurements.
We thank Associated Universities, Inc. and the National Science Foundation for providing summer funding for this research. 
We thank the Atacama Large Millimeter/submillimeter Array for providing the observing hours and calibration for our data. 
AG acknowledges support from the NSF under grants AST 2008101, 2206511, and CAREER 2142300. 
MCS acknowledges financial support from the Royal Society (URF\textbackslash R1\textbackslash 221118) and from the European Research Council via the ERC Synergy Grant ``ECOGAL – Understanding our Galactic ecosystem: from the disk of the Milky Way to the formation sites of stars and planets'' (grant 855130).
The National Radio Astronomy Observatory is a facility of the National Science Foundation operated under cooperative agreement by Associated Universities, Inc.
This paper makes use of the following ALMA data: ADS/JAO.ALMA\#2018.1.00862.S. ALMA is a partnership of ESO (representing its member states), NSF (USA) and NINS (Japan), together with NRC (Canada), MOST and ASIAA (Taiwan), and KASI (Republic of Korea), in cooperation with the Republic of Chile. The Joint ALMA Observatory is operated by ESO, AUI/NRAO and NAOJ.

\vspace{5mm}
\facilities{
    Atacama Compact Array, ALMA 
}

\software{
This research has made use of the following software projects:
    \href{https://astropy.org/}{Astropy} \citep{astropy18},
    \href{https://matplotlib.org/}{Matplotlib} \citep{matplotlib07},
    \href{http://www.numpy.org/}{NumPy},
    \href{https://scipy.org/}{SciPy} \citep{numpy20},
    \href{https://lmfit.github.io/lmfit-py/index.html}{lmfit},
    \href{https://ipython.org/}{IPython} \citep{ipython07},
    \href{https://casa.nrao.edu/}{CASA} \citep{casa07},
    \href{https://github.com/radio-astro-tools/spectral-cube}{spectral-cube},
    \href{https://github.com/radio-astro-tools/radio-beam}{radio-beam},
    \href{https://despotic.readthedocs.io/en/latest/}{DESPOTIC}
    \citep{despotic2014},
    \href{https://aplpy.github.io/}{APLpy} \citep{aplpy2012},
    \href{https://home.strw.leidenuniv.nl/}{RADEX} \citep{radex},
    \href{https://github.com/keflavich/dust_emissivity}{dust\_emissivity},
    and
    the NASA's Astrophysics Data System.
}

\bibliographystyle{aasjournal}
\bibliography{thesis.bib}

\appendix

\begin{figure}[hbp]
    \centering
    \includegraphics[width=0.99\textwidth]{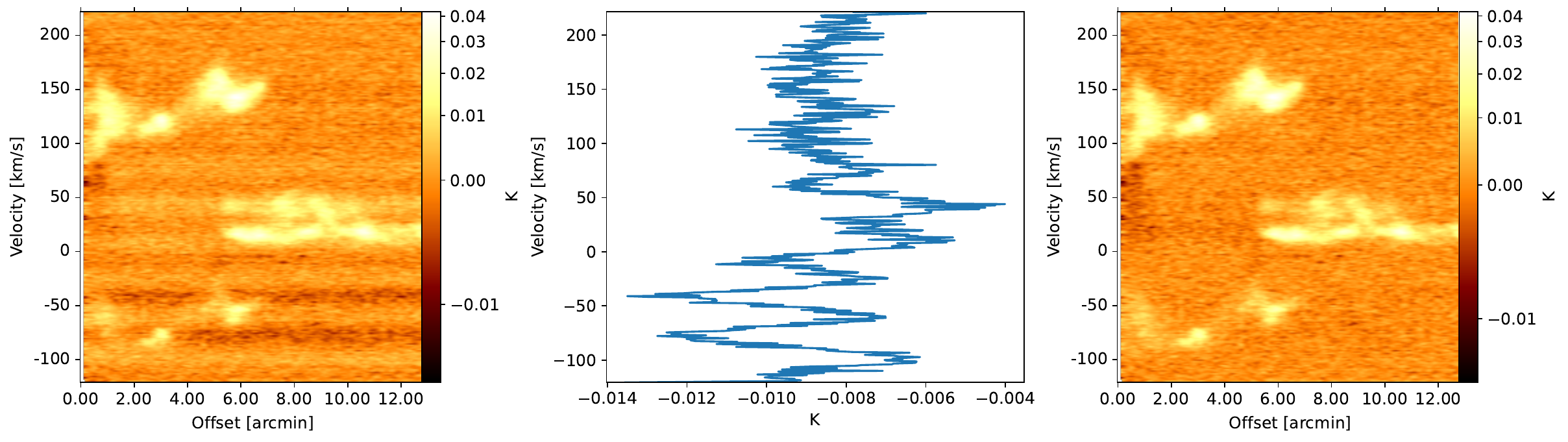}
    \caption{The process of percentile subtraction along the spectral axis, as detailed by Section \ref{sec:ripple}. 
    (Left) PV Diagram of \ce{H2CO} $3_{21}\shortrightarrow2_{20}$ using a raw data cube. Note the spectral ripple. 
    (Middle) 5th percentile spectrum of the raw data. Note that the troughs in the percentile spectrum line up in velocity space with the baseline ripple in the Left panel. 
    (Right) PV Diagram of  \ce{H2CO} $3_{21}\shortrightarrow2_{20}$ after subtracting the 5th percentile spectrum. The baseline ripple has been removed with the baseline centered on \SI{0}{K}.}
    \label{fig:percentile}
\end{figure}

\begin{table}[hbp]
\centering
\caption{Percentile Subtraction}
    \begin{tabular}{|c|c|c|}
    \hline
    Molecule and  & Percentile  & Vertical Shift \\
    Transition & Subtracted & $\mathrm{K}$ \\
     \hline
     \hline
    \ce{HC3N} v=0 $J=24\shortrightarrow23$ & 10 & 0.006 \\
    \ce{H2CO} $J=3_{2,2}\shortrightarrow2_{2,1}$ & 5 & 0.0085 \\
    \ce{H2CO} $J=3_{0,3}\shortrightarrow2_{0,2}$ & 1 & 0.011 \\
    SiO v=0 $J=5\shortrightarrow4$ & 5 & 0.0124 \\
    \ce{C^18O} $J=2\shortrightarrow1$ & 1 & 0.0215 \\
    \hline
    \end{tabular}
\label{tbl:persub}
\end{table}

\begin{figure}
    \centering
    \includegraphics[width=1\textwidth]{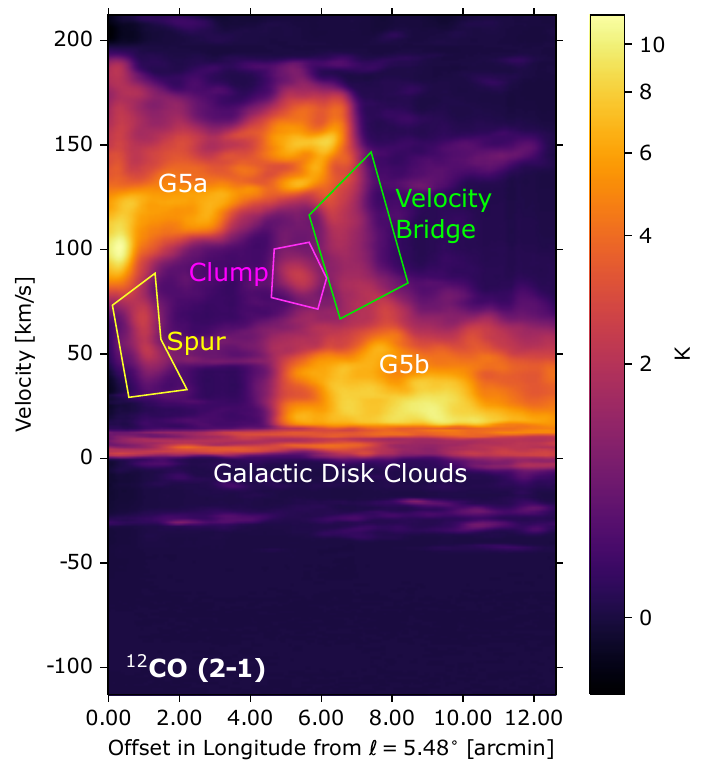}
    \caption{Position-Velocity diagram of Field 2 in \ce{^12CO} $J=2\shortrightarrow1$, averaged over Galactic latitude and taken horizontally across the field with a width of \SI{2}{\arcmin}. This is the same as Figure \ref{fig:pvdiagram}, but with the features detailed in Section \ref{sec:pvdiagrams} labelled.}
    \label{fig:labelled_pv}
\end{figure}

\begin{figure}
    \centering
    \includegraphics[width=0.7\textwidth]{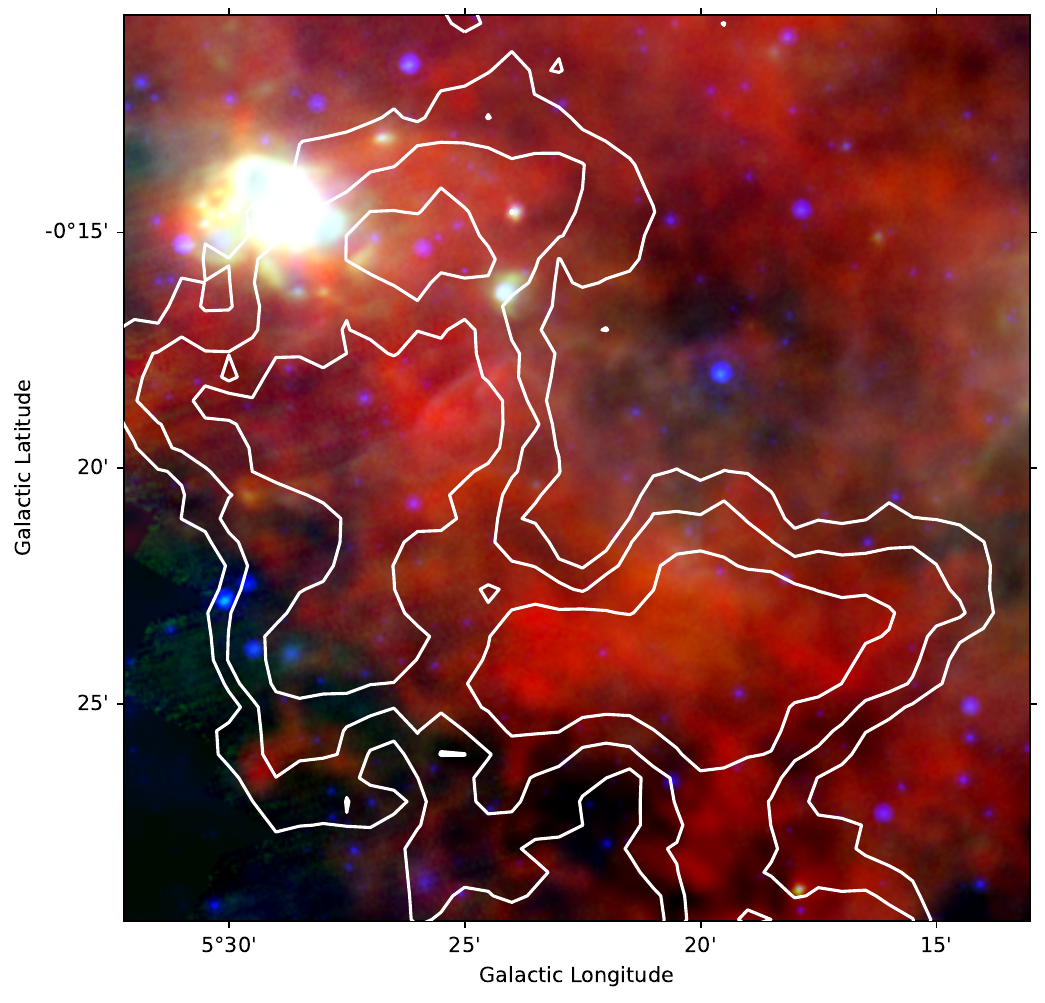}
    \caption{Three color image of Herschel SPIRE \SI{250}{\mu m}, PACS \SI{70}{\mu m}, and MIPSGAL \SI{24}{\mu m} overlaid with \ce{NH3} (3,3) integrated intensity contours [6.8, 11.0, 17.8] from the Mopra HOPS Survey \citep{hops3, hops1, hops2}.}
    \label{fig:dustshape}
\end{figure}

\end{document}